\documentclass[11pt]{article}

\usepackage[final]{acl}

\usepackage{times}
\usepackage{latexsym}
\usepackage{amsmath}

\usepackage[T1]{fontenc}

\usepackage[utf8]{inputenc}

\usepackage{microtype}

\usepackage{inconsolata}
\usepackage{array}

\usepackage{graphicx}

\usepackage{booktabs}

\usepackage{subcaption}

\usepackage{xcolor}
\usepackage{tcolorbox}

\usepackage{xurl}   %
\usepackage{hyperref}

\newcommand{\regular}{non-review}

\definecolor{maria-color}{HTML}{7881F2}

\usepackage{xcolor}
\usepackage{colortbl}
\usepackage{booktabs}
\definecolor{tableblue}{RGB}{66, 133, 244}
\definecolor{tablered}{RGB}{219, 68, 55}

\title{LLM-Generated or Human-Written? \\ Comparing Review and Non-Review Papers on ArXiv}

\author{Yanai Elazar \\
  Bar-Ilan University \\  
  \texttt{yanaiela@gmail.com} 
\And
  Maria Antoniak \\
  University of Colorado Boulder \\
  \texttt{maria.antoniak@colorado.edu} }

\begin{document}
\maketitle
\begin{abstract}
ArXiv recently prohibited the upload of unpublished review papers to its servers in the Computer Science domain, citing a high prevalence of LLM-generated content in these categories. However, this decision was not accompanied by quantitative evidence. In this work, we investigate this claim by measuring the proportion of LLM-generated content in review vs. non-review research papers in recent years.
Using two high-quality detection methods, we find a substantial increase in LLM-generated content across both review and \regular{} papers, with a higher prevalence in review papers. 
However, when considering the number of LLM-generated papers published in each category, the estimates of \regular{} LLM-generated papers are almost six times higher.
Furthermore, we find that this policy will affect papers in certain domains far more than others, with the CS subdiscipline Computers \& Society potentially facing cuts of 50\%.
Our analysis provides an evidence-based framework for evaluating such policy decisions, and we release our code to facilitate future investigations at: \url{https://github.com/yanaiela/llm-review-arxiv}.
\end{abstract}

\section{Introduction}

On October 31st, 2025, arXiv announced a policy change\footnote{While the blog post explicitly states that this is not a technical policy change, it is a de-facto change.} prohibiting the upload of review,\footnote{Throughout this paper, we use the term \textit{review papers} to refer to all three types of papers covered by arXiv's policy, which refers to ``review, survey, and position'' papers.} survey, and position papers to their Computer Science (CS) servers~\cite{arxiv-blog}.
According to the blog post, this decision was motivated by an observed increase in LLM-generated content in review papers in the CS subdomain over recent years. Since papers uploaded to arXiv undergo moderation by volunteer human experts, overburdened with other obligations, the administrators decided to restrict further approvals of unpublished review papers altogether.
The blog post did not provide quantitative evidence or data to support this claim. In this paper, we address two central questions:
\emph{Do review papers contain higher proportions of LLM-generated content than regular research papers? What downstream effects would be created by this policy?}

\begin{figure}[t]
\centering
\includegraphics[width=0.85\columnwidth]{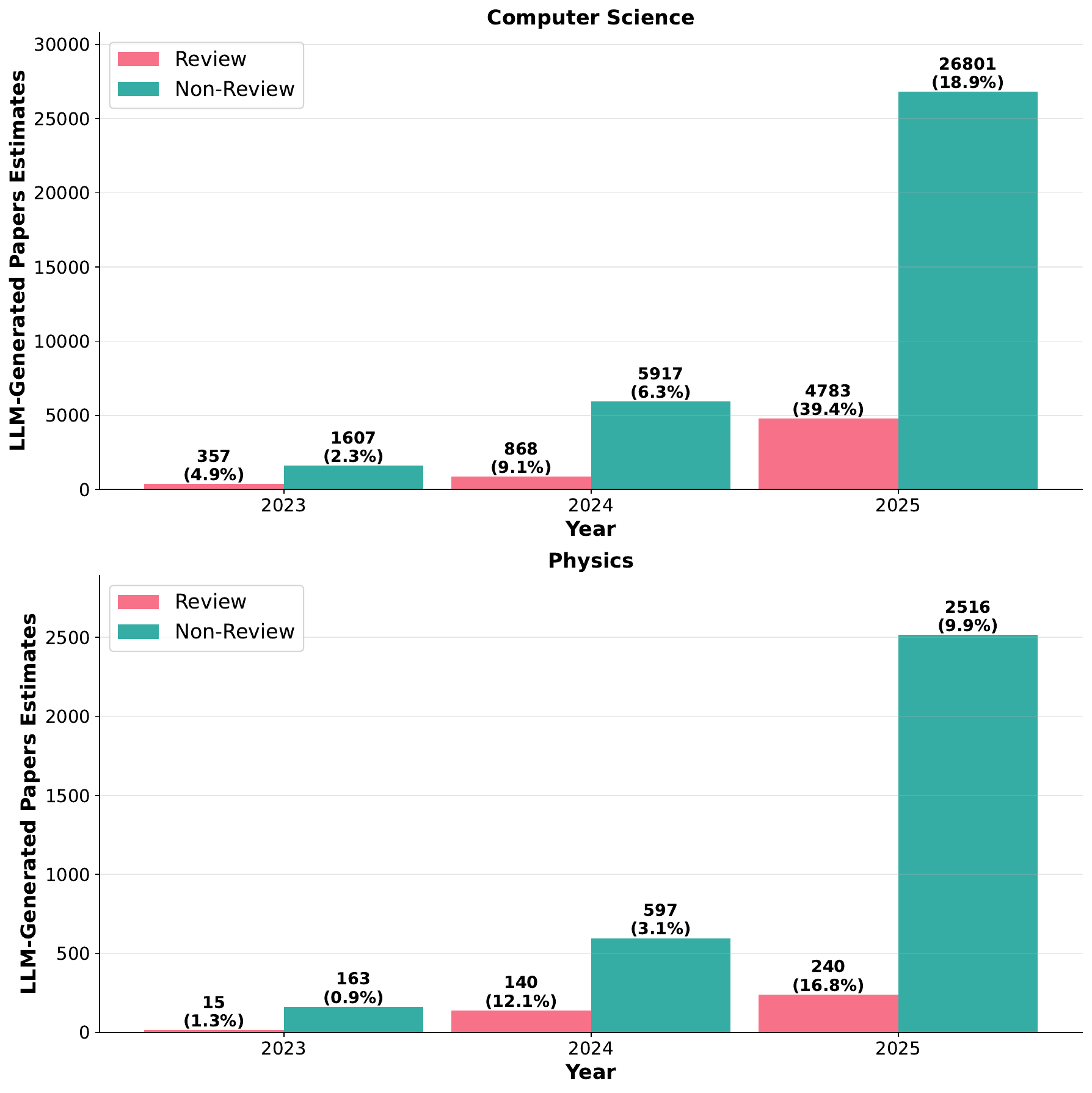}
\caption{LLM-generated paper estimates using Pangram detection for review vs. non-review papers. While review papers show higher percentages of LLM-generated content (in parentheses), the absolute number of non-review papers detected as LLM-generated is substantially larger in all domains, suggesting that restricting review papers alone may not address the core concern about LLM-generated content on arXiv.}
\label{fig:pangram-estimates-cs-physics}
\end{figure}

The proliferation of advanced AI writing assistants, particularly large language models (LLMs), has enabled researchers to incorporate these tools into their scientific workflows \citep{nejjar2025llms,zhang2025advancing,eger2025transforming,si2025can}. In the extreme case, papers may be fully written by LLMs, including literature reviews and experimental design.\footnote{Several instances of fully LLM-generated papers have been reported in the literature~\citep{lu2024ai}.}
While we have preliminary evidence for fully LLM-generated papers receiving scores sufficient for workshop publication \citep{yamada2025ai}, a central concern is whether LLM-generated papers maintain the same \textit{standards} as human-written papers.
The ease with which papers can be generated, and the increasing rates of paper submissions, especially about AI topics, provoke concerns about overwhelm for arXiv moderators.

These are legitimate concerns, but the focus on review papers makes several assumptions that may or may not be accurate or helpful.
First is the central assumption that review papers are more likely to be generated than non-review papers.
Second is a set of epistemic assumptions about what constitutes a review paper and what purposes review papers serve in a research community, reflecting disciplinary biases (not shared across all CS subdisciplines) that elevate quantitative over qualitative methods and evidence.

In this paper, we quantitatively evaluate this new policy. 
We evaluate the ratios of review papers that are LLM-generated vs. \regular{} papers, the temporal trends of such ratios, and how these trends vary across fields.
Our main analysis classification and detection pipeline uses only paper titles and abstracts (not full text), reflecting the data used by arXiv to make the initial decision. We then run an additional analysis on a subset of papers for which we run the analysis on the full text, where we validate the consistency of our results.
As such, we employ a two-stage methodology. First, we develop an LLM-based classifier to distinguish review papers from regular research papers, achieving high accuracy (92.0\% F1) on manually annotated data. Second, we apply two LLM-generation detection methods that estimate the fraction of LLM-generated content in a corpus to measure LLM prevalence. We analyze a large dataset of arXiv papers from different domains to measure these rates.

Our findings indicate that: (1) review papers contain higher rates of LLM-generated content than \regular{} papers in CS, with adjusted cohort-level estimates of 21.4\% for review papers versus 14.0\% for \regular{} papers; (2) LLM-generated content has increased dramatically in both categories following ChatGPT's release in November 2022, with review papers showing an increase from 12.9\% in 2023 to 28.2\% in 2025 and \regular{} papers from 6.2\% to 18.9\%; and (3) this pattern persists across all examined disciplines (Computer Science, Physics, Mathematics, Statistics) and shows an accelerating trend since 2023.

Crucially, when considering the number of papers published for each type, we estimate that in CS, there are almost six times more \regular{} papers generated by LLMs than review papers (Figure \ref{fig:pangram-estimates-cs-physics}). This result throws into doubt arXiv's decision to restrict the upload of unpublished review papers to their CS servers.
We also find that the policy will affect certain disciplines and topics differently; topics like education and safety have much higher rates of review papers, and Computers \& Society papers could be censored by 50\%, while Computer Vision will only lose 3\% of its papers.
We hope this work provides an evidence-based perspective on arXiv's policy decision while fostering broader discussion about the long-term implications of LLM-generated content in scientific publishing.

\section{Background} \label{sec:related}

\paragraph{A Brief History of ArXiv.}

Arxiv originally started as an automated email server before converting into a website. As its usage and submissions grew, arXiv faced challenges related to software scale and content moderation \citep{han2025arxiv}.\footnote{\url{https://arxiv.org/stats/monthly_submissions}}
While arXiv has historically relied on volunteer moderators to ensure basic quality standards, the scale and nature of submissions have evolved significantly, and as such, automatic classifiers were introduced to assist with moderation. The October 2025 policy restricting review papers in CS represents a notable departure from arXiv's traditional open policy.\footnote{\url{https://info.arxiv.org/help/policies/}}

\paragraph{LLM-Generated Text Detection.}
The detection of LLM-generated text has become an active research area following the release of powerful language models~\citep{gpt3,gpt4}. Early approaches relied on statistical features such as perplexity and token probability distributions~\citep{mitchell2023detectgpt}. More recent work has explored zero-shot detection methods~\citep{mitchell2023detectgpt}, watermarking techniques~\citep{kirchenbauer2023watermark}, and distributional estimation approaches that measure the fraction of AI content in document collections rather than classifying individual texts~\citep{sadasivan2024aidetection}.
Beyond the design of accurate detection models, there is the question about the right task definition. For instance, \citet{ghosal2023survey} argues that binary AI-versus-human classifications are insufficient, proposing instead to estimate the percentage of LLM-generated content within a document, with new definitions of such tasks and proposing novel methods \citep{wang2023seqxgpt,wang2025real,thai2025editlens}.

The \textit{Alpha} estimator method~\citep{liang2024monitoring} represents a shift from instance-level to population-level detection. Rather than attempting to classify individual documents with high confidence, it uses maximum likelihood estimation to measure the fraction of LLM-generated content across an entire corpus. This approach is particularly well-suited for our research question, which focuses on comparing aggregate prevalence rates between paper categories.

\paragraph{LLMs in Academic Writing.}
Recent studies have documented the increasing use of LLM writing assistants and \textit{agents} in academic contexts~\citep{liang2024monitoring}. 
Concerns have been raised about the impact on scientific integrity, originality, and the peer review process~\citep{lin2025chatgptlinguisticequalizerquantifying,latona2024aireviewlotterywidespread}. 
Several instances of fully or largely LLM-generated papers have been identified in published literature~\citep{dupre2023researchers}, raising questions about quality control mechanisms. 

Some work has examined disciplinary differences in AI adoption, finding that fields with greater exposure to AI technology show earlier and higher adoption rates~\citep{liao2025llms}. However, to our knowledge, no prior work has systematically compared generated content between review papers and regular research papers or provided quantitative evidence for preprint server policies.

\section{Methods} \label{sec:method}

\subsection{LLM-Generated Content Detection}

We focus on estimating population-level trends (e.g., the proportion of LLM-generated content among all review papers posted on arXiv in 2025) and comparing these trends across groups (e.g., review vs. non-review, 2023 vs. 2025). While this is a challenging task, and we do not expect perfect accuracy, we can estimate and compare aggregate statistics over populations.\footnote{Assuming that any systematic biases in the classifier affect both groups similarly.}

We employ two complementary detection methods: (1) the Alpha estimator~\citep{liang2024monitoring}, which provides a group-level estimate of the fraction of LLM-generated content based on the occurrence of adjectives used in a document, and (2) a commercial LLM-detector called Pangram~\citep{emi2024technical},\footnote{\url{https://www.pangram.com/}} a transformer-based \citep{transformers} classifier trained to distinguish LLM-generated text from human-written content across multiple domains and language models.
These methods are complementary: they were tuned on different datasets, they operate on different inputs, and they model the problem differently.

\paragraph{Evaluation.}

Our estimators were previously evaluated and shown to be highly accurate (below 2.4\%  error rates for Alpha \citep{liang2024monitoring}, and below 1\% for Pangram \citep{emi2024technical}).
While errors during the classification of papers post-LLM era are practically impossible for us to evaluate as we do not have access to the ground truth, we evaluate them on papers from the pre-LLM era to measure false positive errors.
We test both Alpha and Pangram on CS papers from 2020-2022.
We find that while Pangram did not predict any false positives, the Alpha method predicts a non-trivial amount of LLM-generated papers in the pre-LLM era: 13.1\% and 1.9\% for review and \regular{} papers, respectively. As such, we employ a correction to the Alpha estimates to correct for such bias.
While the pre-LLM period may still include usage of writing assistance (e.g., grammar/style editing tools, by tools such as Grammarly \footnote{\url{https://www.grammarly.com/}}), such tools typically minimally edited texts locally, and do not imply LLM text generation; we therefore treat the pre-LLM signal primarily as detector bias and correct for it explicitly.

\paragraph{Alpha Estimates.}

We employ a distributional detection approach using the Alpha estimator method~\citep{liang2024monitoring}, which uses maximum likelihood estimation (MLE) to measure the fraction $\alpha$ of LLM-generated or LLM-modified content across an entire dataset (group).
For each group we compute Alpha, representing the estimated ratio of LLM-generated papers, and the 95\% confidence interval of the estimation.

\paragraph{Correction.}
Due to non-trivial rates of false positives in the Alpha estimation in the pre-LLM era, we use these values to calibrate our post-LLM estimates via the Rogan-Gladen adjustment \citep{rogan1978estimating}. This adjustment takes into account false positive and false negative rates to correct apparent (inaccurate) prevalence measurements.
We use the detection rates of the pre-LLM estimates for Alpha per group: review and \regular{} papers, to correct every group estimate. Additional details on the Rogan-Gladen adjustment as well as our implementation and assumptions are provided in Appendix \ref{app:rogan-gladen-adjustment}.
For readability, we center the main analysis on corrected estimates because the uncorrected Alpha values are known to be biased upward in our validation.

\paragraph{Pangram Estimates.}

We use the Pangram API \citep{emi2024technical} 
and treat ``Likely AI'', ``Possibly AI'', and ``Highly Likely AI'' labels as LLM generated papers. ``Unlikely AI'' classifications are treated as human written.
For each group, we compute the proportions of LLM-generated papers, and the standard error of the mean of the distribution. %

\subsection{Review Paper Classification}
\label{subsec:review-classification}
The arXiv policy relies on distinguishing between ``review'' vs. non-review papers. 
But what is a ``review'' paper? 
While this question might seem obvious, such taxonomies are neither universal nor neutral; 
different fields can have different epistemic cultures and ``machineries of knowledge construction'' \citep{cetina1999epistemic}, and interdisciplinary work often resists such categorizations \citep{clement2016methodology}.

ArXiv's CS corpus spans communities with different epistemic cultures and publication norms, from established subfields like Programming Languages (CS.PL) to interdisciplinary areas like Computation \& Language (CS.CL) and Computers \& Society (CS.CY).
The latter two include papers from the digital humanities (DH), computational social science, and fairness, accountability, and transparency research (FAccT), where community norms are known to clash \citep{laufer2022four}.

While we do not know exactly how arXiv moderators categorize review papers, we approximate what such a process might look like.%
\footnote{However, any such classifier will necessarily impose one epistemic culture's categories onto papers from others.}

\paragraph{What are Review Papers?}
\textit{Review papers}, which we consider to include \textit{survey papers},
often aim to synthesize and summarize a body of literature, identify trends over time, provide recommendations and guidelines for best practices, and introduce newcomers to a field.
Structurally, review papers tend to emphasize literature coverage over novel experimental contributions, though high-quality reviews require substantial domain expertise.
Some reviews offer novel insights through meta-analysis, and many different kinds of review papers exist, ranging from critical and mixed-method reviews to standardized Cochrane reports \citep{grant2009typology}.
\textit{Position papers} (included in arXiv's ban) often function like editorials; they offer opinionated perspectives on current issues in a discipline and tend to rely more heavily on literature review than experimental results. 
However, these are generalizations based on traditional CS disciplines like machine learning, and we emphasize that this definitional task represents epistemic choices that are dependent on discipline.

\paragraph{Review vs. Non-Review Dataset.}
To test the feasibility of the task and to evaluate the performance of our review paper classifier, we create a manually annotated dataset sampled from the arXiv Computer Science category across different years.
We hand-annotate a set of 200 papers as review or non-review papers.
These papers were sampled such that 100 of them had keywords related to reviews\footnote{‘review’, ‘survey’, ‘overview’, ‘primer’, ‘systematic review’, and ‘literature review’.} in their title or abstract, while the other 100 papers included ``experimental'' keywords.\footnote{‘results’, ‘experimental’, ‘empirical’, ‘evaluation’, and ‘dataset’.}
Two of the paper's authors, both with NLP and CS expertise, discussed examples and definitions and then independently labeled each paper based on its title and abstract, achieving strong agreement (Cohen's $\kappa$ = 0.85).
We use 60 papers for evaluation, and 140 papers as a test set.
We use one annotator's labels as the ground truth for evaluation purposes.

\paragraph{Evaluation.}
We evaluate the following LLMs for classifying review vs. non-review: Llama 3.3 \citep{llama3}, Gemma 2 \citep{team2024gemma}, GPT-OSS \citep{agarwal2025gpt}, and GPT-4o-mini \citep{hurst2024gpt} on our validation set, achieving 95.4, 95.6, 92.6, and 95.6 F1 scores, respectively. We select Gemma 2 as our classification model as it achieved the best performance on the validation set, and we can run it locally.
We report the results on the held-out test set in Table~\ref{tab:classification-results}, as well as the prompt used for all models in Appendix \ref{app:review}.

\subsection{Quantifying LLM-Generated Content}

We apply the review paper classifier to a large corpus of arXiv papers, classifying them into review and non-review papers. We then apply the AI content detector to both groups\footnote{Using Rogan-Gladen adjustment to correct for the false positive of the Alpha method, and simply computing the ratio of LLM-generated papers using Pangram.}
in each category and perform statistical significance tests: 95\% confidence intervals over $30K$ bootstrapped samples for Alpha following \citet{liang2024monitoring}, and standard error of the mean for Pangram. 
Our manual ground-truth annotation is CS-only by design, because the arXiv policy under study currently targets CS; we then apply the same title/abstract-based classifier across other domains for comparative context rather than policy-specific labeling claims about those domains.
We use the entire dataset to compute the estimates using alpha, while using a sample of 100 papers from every month with Pangram, to save costs, for a total of 3,600 per domain for the post-LLM period.

\begin{figure*}[t]
\centering
\begin{subfigure}[b]{0.7\textwidth}
\centering
\includegraphics[width=\textwidth]{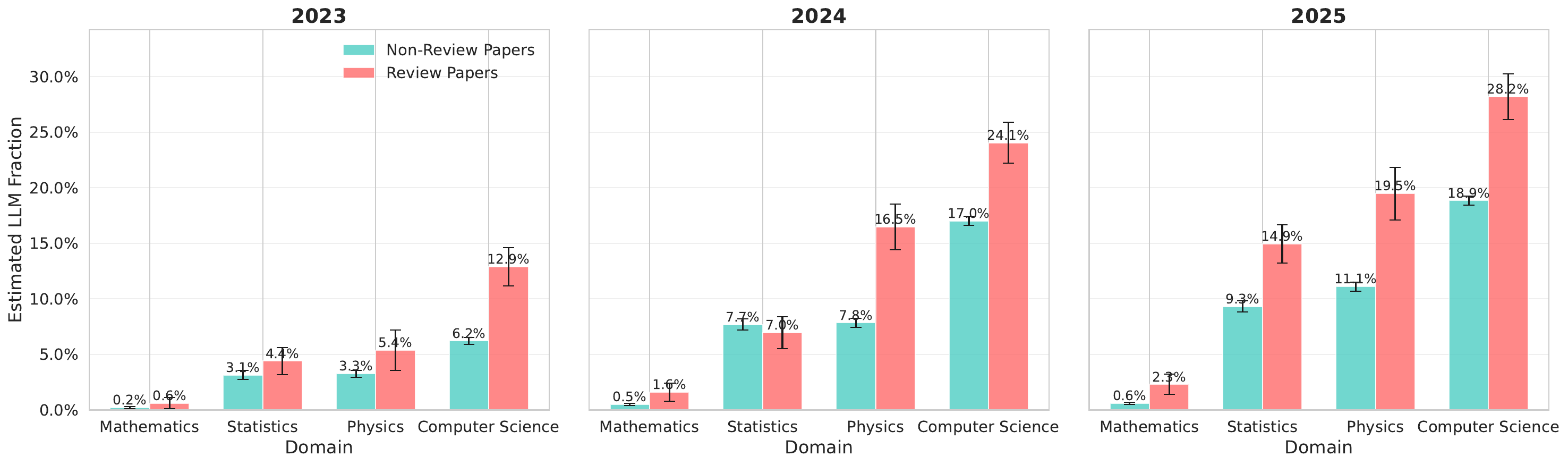}
\caption{Alpha estimates.}
\label{fig:adjusted-alpha-category-year-alpha}
\end{subfigure}

\begin{subfigure}[b]{0.7\textwidth}
\centering
\includegraphics[width=\textwidth]{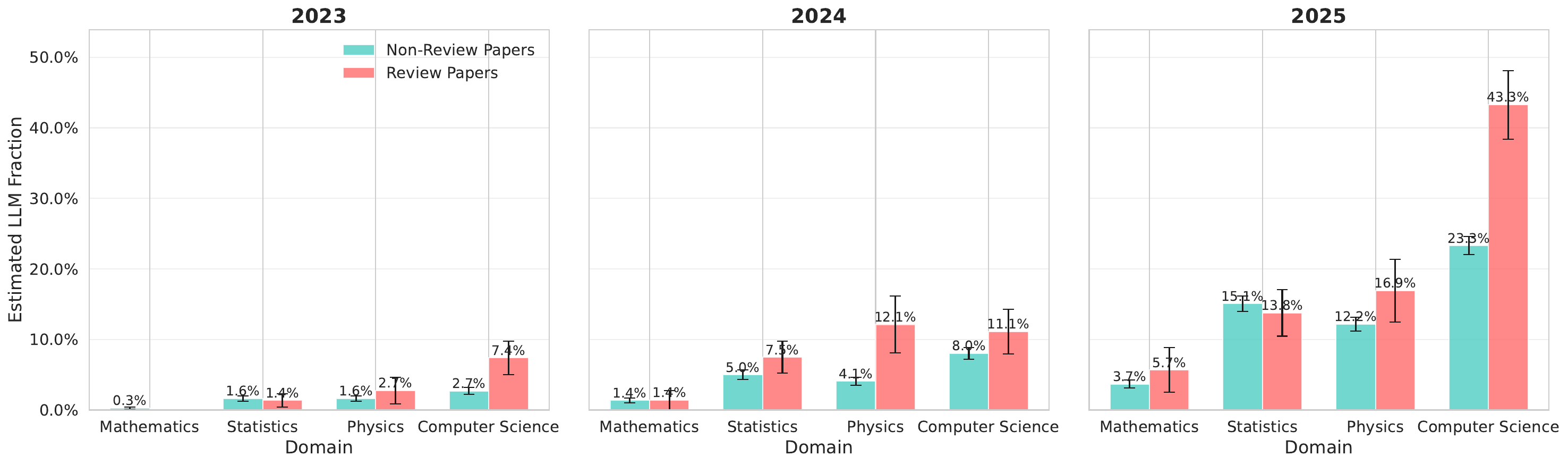}
\caption{Pangram ratios.}
\label{fig:adjusted-alpha-category-year-pangram}
\end{subfigure}
\caption{LLM-generated papers estimates using (a) Alpha estimates and (b) Pangram detection ratios on the \textit{arxiv-domains} dataset. Both methods reveal temporal patterns in LLM-adoption across domains, and years.}
\label{fig:adjusted-alpha-category-year}
\end{figure*}

\subsection{Datasets}
We use an official arXiv dataset released on Kaggle by Cornell University,\footnote{\url{https://www.kaggle.com/datasets/Cornell-University/arxiv}, sampled on December 15th, 2025.} which contains metadata, abstracts and links to the PDF files for posted papers, in English.
We focus on papers from Computer Science, Mathematics, Statistics, and Physics, published between 2020 and 2025, allowing us to evaluate and correct the estimates using the pre-LLM (2020-2022) era and to perform our main analysis on the post-LLM (2023-2025) era.

Our first dataset, \textit{arxiv-domains}, consists of four popular domains (CS, Mathematics, Statistics, and Physics), with up to $36,000$ papers from arXiv (500 per month, for every month between 2020-2025 $=500*12*6$), including metadata: titles, abstracts, publication dates, and categories. Overall, we collect $124,461$ papers for \textit{arxiv-domains}.
In addition, we collect a dataset of CS subcategories,\footnote{The ten most common CS subcategories: \textit{Artificial Intelligence},
\textit{Computation and Language},
\textit{Cryptography and Security},
\textit{Computer Vision and Pattern Recognition},
\textit{Computers and Society},
\textit{Human-Computer Interaction},
\textit{Information Retrieval},
\textit{Machine Learning},
\textit{Robotics}, and
\textit{Software Engineering}} which we name \textit{cs-subcategories}, sampling up to 500 papers per month per subcategory, up to 36,000 per subcategory.
Overall, we collect $138,244$ papers for \textit{cs-subcategories}. We provide more detailed statistics in Appendix \ref{app:dataset-stats}.
Unless noted otherwise, we use titles and abstracts for classification and detection, and we discuss this decision in \S\ref{sec:discussion}. We also experiment with analyzing the full texts in \S \ref{subsec:full-text-analysis}.

\section{Results} \label{sec:results}

\subsection{Rates vary dramatically by discipline}

Figures \ref{fig:adjusted-alpha-category-year}-\ref{fig:adjusted-alpha-cs-subcat}, as well as \ref{fig:adjusted-alpha-category}-\ref{fig:adjusted-alpha-cs-subcat-year} in the Appendix, show that review papers tend to have higher rates of LLM-generated content than \regular{} papers in CS. This trend holds across domains and when using both the Alpha estimates and Pangram. We elaborate on these results in Appendix \ref{app:rates-expanded}.

\subsection{Generated papers are increasing}
\label{subsection:upward-trends}

We then examine how these patterns have shifted over time.
Figure~\ref{fig:adjusted-alpha-category-year} shows the LLM-generated content estimates by year.
We observe a dramatic increase in LLM-generated content since the release of ChatGPT between 2023-2025, consistent with the findings of \citet{liangmapping}.

In CS, the Alpha in review papers increases from $12.9\%$ in 2023 to $28.2\%$ in 2025. For \regular{} papers, the increase was from $6.2\%$ in 2023 to $18.9\%$ in 2025. The gap between review and \regular{} papers has remained substantial and significant in all years.
The Pangram results show similar, but more dramatic trends. In CS, the percentage of AI detected papers grew from $7.4\%$ in 2023 to $43.3\%$ in 2025 in review papers, compared with $2.7\%$ to $23.3\%$ in \regular{} papers.

\subsection{Rates vary dramatically by subcategory}
\label{subsec:subcategories}

We conduct a fine-grained analysis examining AI adoption patterns across different CS subcategories. 
Overall, we find large variations in review paper and AI-generated paper rates.
Table \ref{table:cs-categories} shows that Computers \& Society has a far higher rate of review papers than other subcategories (49.2\%, compared to the next highest category, Software Engineering, with 23.4\%). 
We find a high Pearson correlation ($\rho=0.886$) between the review and LLM-generated paper rates.

\begin{figure}[t]
\centering
\begin{subfigure}[b]{0.85\columnwidth}
\centering
\includegraphics[width=\textwidth]{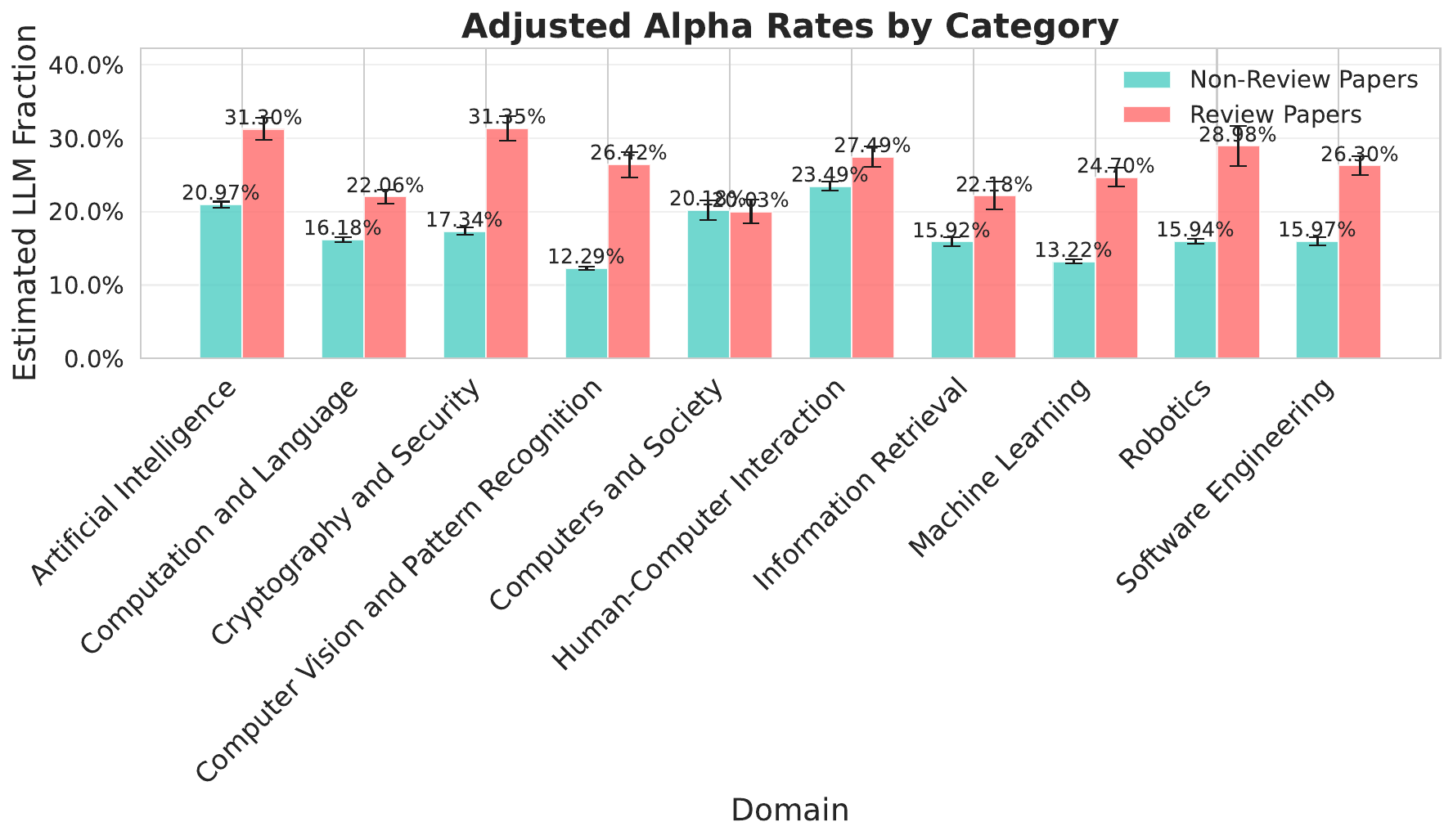}
\caption{Alpha estimator method}
\label{fig:adjusted-alpha-cs-subcat-alpha}
\end{subfigure}

\begin{subfigure}[b]{0.85\columnwidth}
\centering
\includegraphics[width=\textwidth]{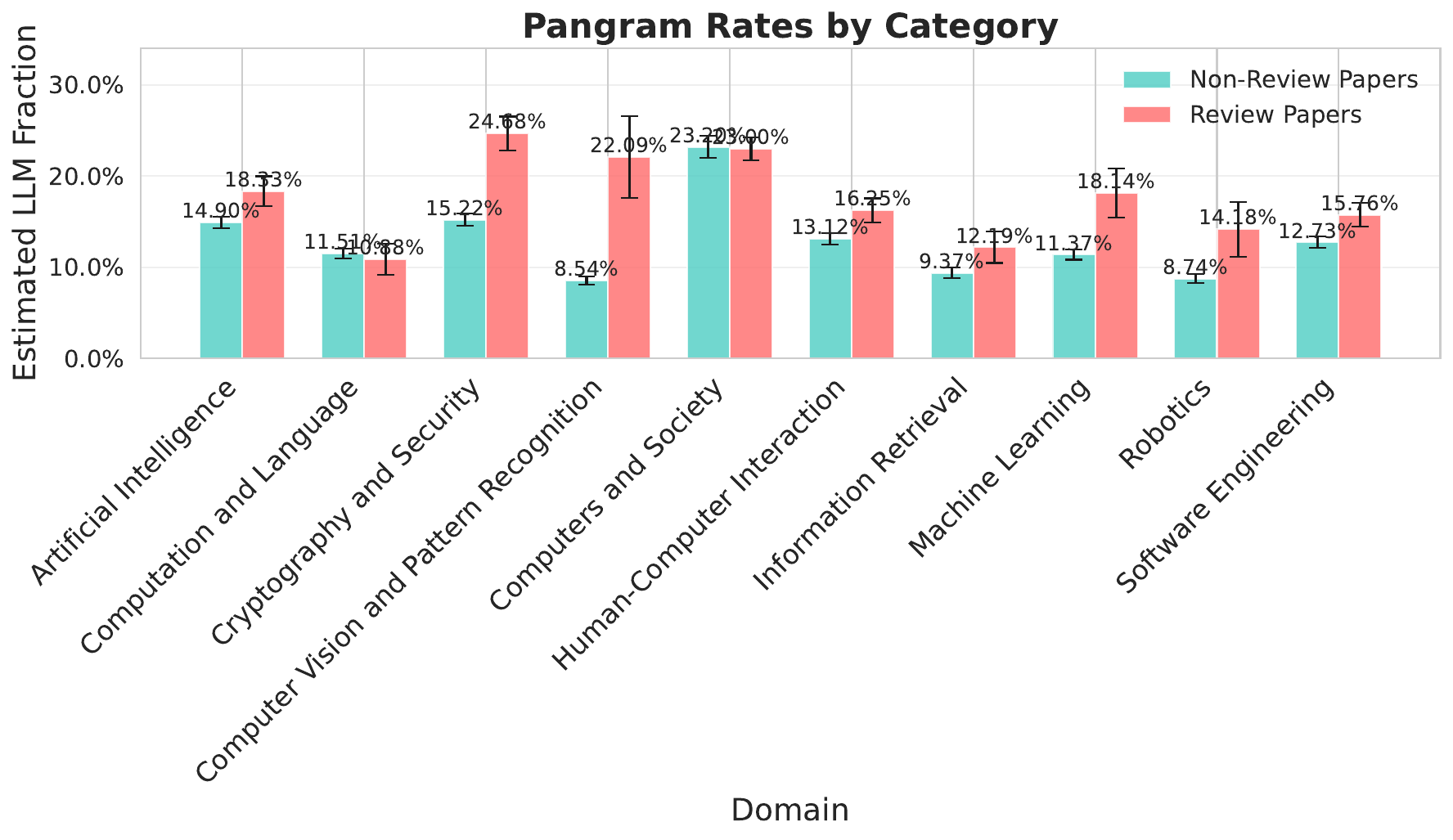}
\caption{Pangram detection method}
\label{fig:adjusted-alpha-cs-subcat-pangram}
\end{subfigure}
\caption{LLM-generated papers estimates by subcategories and paper type in \textit{cs-subcategories} using (a) Alpha estimates and (b) Pangram detection ratios. The review versus non-review paper gap (or lack thereof) varies considerably across subcategories, suggesting that field-specific norms influence AI adoption patterns.}
\label{fig:adjusted-alpha-cs-subcat}
\end{figure}

Figure~\ref{fig:adjusted-alpha-cs-subcat} presents the Alpha and Pangram estimates for review and \regular{} papers across major CS subcategories. 
Overall, the predicted rates for review papers are always higher or equivalent to the predicted rates for non-review papers.
The analysis reveals some variance within CS, with some subcategories showing much higher AI adoption rates than others.
For example, Computation and Language (the NLP subcategory) shows the lowest use of AI in review papers (22.06\% and 10.88\% using the Alpha and Pangram estimates, respectively).
On the other hand, Cryptography and Security shows the highest adoption rates for review papers, with estimates of 31.35\% and 24.68\% using Alpha and Pangram.\footnote{This result is unintuitive, and we leave it as an open question for future work.}

Figure~\ref{fig:adjusted-alpha-cs-subcat-year} in the Appendix further breaks down these patterns over time. %
These subcategory-level findings have important policy implications. ArXiv's blanket restriction on CS review papers treats all CS subcategories uniformly, despite dramatic differences in AI adoption rates and community standards within CS. 

\begin{table}[t]
\footnotesize
\centering
\begin{tabular}{lcc}
\toprule
\textbf{Category} & \textbf{Review Rate} & \textbf{AI Rate} \\
\midrule
Computers \& Society & \cellcolor{blue!50} 49.2\% & \cellcolor{red!40} 20.2\% \\
Software Engineering & \cellcolor{blue!24} 23.4\% & \cellcolor{red!21} 10.6\% \\
HCI & \cellcolor{blue!23} 23.3\% & \cellcolor{red!23} 11.3\% \\
AI & \cellcolor{blue!16} 15.8\% & \cellcolor{red!26} 13.0\% \\
Cryptography \& Security & \cellcolor{blue!15} 15.2\% & \cellcolor{red!28} 13.8\% \\
Information Retrieval & \cellcolor{blue!13} 13.4\% & \cellcolor{red!16} 7.8\% \\
Computation \& Language & \cellcolor{blue!9} 9.4\% & \cellcolor{red!18} 8.8\% \\
Machine Learning & \cellcolor{blue!6} 5.7\% & \cellcolor{red!20} 9.9\% \\
Robotics & \cellcolor{blue!4} 3.7\% & \cellcolor{red!13} 6.7\% \\
Computer Vision & \cellcolor{blue!2} 2.4\% & \cellcolor{red!14} 7.0\% \\
\bottomrule
\end{tabular}
\caption{Review vs.\ AI-generated rates across CS in the \textit{cs-subcategories} dataset. The Pearson correlation between the review and AI columns is high ($\rho=0.886$).}
\label{table:cs-categories}
\end{table}

\subsection{Who is affected by the policy change?}

Using the OpenAlex API,\footnote{OpenAlex is an online library: \href{https://openalex.org/}{https://openalex.org/}.} we explore what kinds of papers and authors are more affected by the arXiv policy change. 
For each paper in the \textit{cs-subcategories} dataset (\char`\~138K papers), we use the paper's DOI to collect the paper's topics, keywords, citation count, and authors, as well as each author's affiliation, h-index, number of works, and citations count, and each institution's country and h-index.
We identify 82\% of the papers (with reasonable coverage across the categories, ranging 69-86\%), and of these, 98\% have at least one resolved author and 100\% had identifiable topics.
A manual audit by one of the authors of 20 papers found two papers with topic errors, while an audit of 20 authors found 9 errors (most merged profiles of authors with ambiguous names).
We therefore include author-focused results only in the Appendix and with strong caveats.
See Appendix ~\ref{appendix-openalex} for more details.

\begin{table}[t]
\footnotesize
\centering
\begin{tabular}{lcc}
\toprule
\textbf{Topic} & \textbf{Review Rate} & \textbf{AI Rate} \\
\midrule
Safety Research & \cellcolor{blue!50} 59.7 & \cellcolor{red!30} 15.9 \\
Health Informatics & \cellcolor{blue!39} 45.5 & \cellcolor{red!50} 23.6 \\
Education & \cellcolor{blue!31} 33.8 & \cellcolor{red!32} 16.5 \\
Sociology \& Poli. Sci. & \cellcolor{blue!31} 33.2 & \cellcolor{red!18} 11.3 \\
Management Info. Systems & \cellcolor{blue!30} 31.2 & \cellcolor{red!35} 18.1 \\
Ocean Engineering & \cellcolor{blue!11} 4.5 & \cellcolor{red!15} 9.8 \\
Aerospace Engineering & \cellcolor{blue!11} 4.1 & \cellcolor{red!12} 8.7 \\
Control \& Systems Enginer. & \cellcolor{blue!10} 3.8 & \cellcolor{red!13} 9.4 \\
Biomedical Engineering & \cellcolor{blue!10} 3.4 & \cellcolor{red!15} 9.9 \\
Computational Mechanics & \cellcolor{blue!10} 2.6 & \cellcolor{red!10} 7.9 \\
\bottomrule
\end{tabular}
\caption{The ten OpenAlex topics with the highest and lowest predicted review paper rates in \textit{cs-subcategories}. Only topics with at least 100 identified papers are shown.}
\label{tabel:openalex-topics}
\end{table}

We find that papers associated with topics like ``safety research,'' ``sociology \& political science,'' and ``education'' are more likely to be review papers, and while these papers also tend to higher Pangram-predicted rates of generated text, their ranking does not exactly match the ranking of rate of reviews (see Table~\ref{tabel:openalex-topics}).
These results confirm the category patterns from Table~\ref{table:cs-categories} but add more detail. 
We note that one of the topics, ``health informatics'' contains many non-healthcare related papers that are focused on education and related social topics.

\subsection{Do our results change for full texts?}
\label{subsec:full-text-analysis}

\begin{figure}[t]
\centering
\includegraphics[width=0.85\columnwidth]{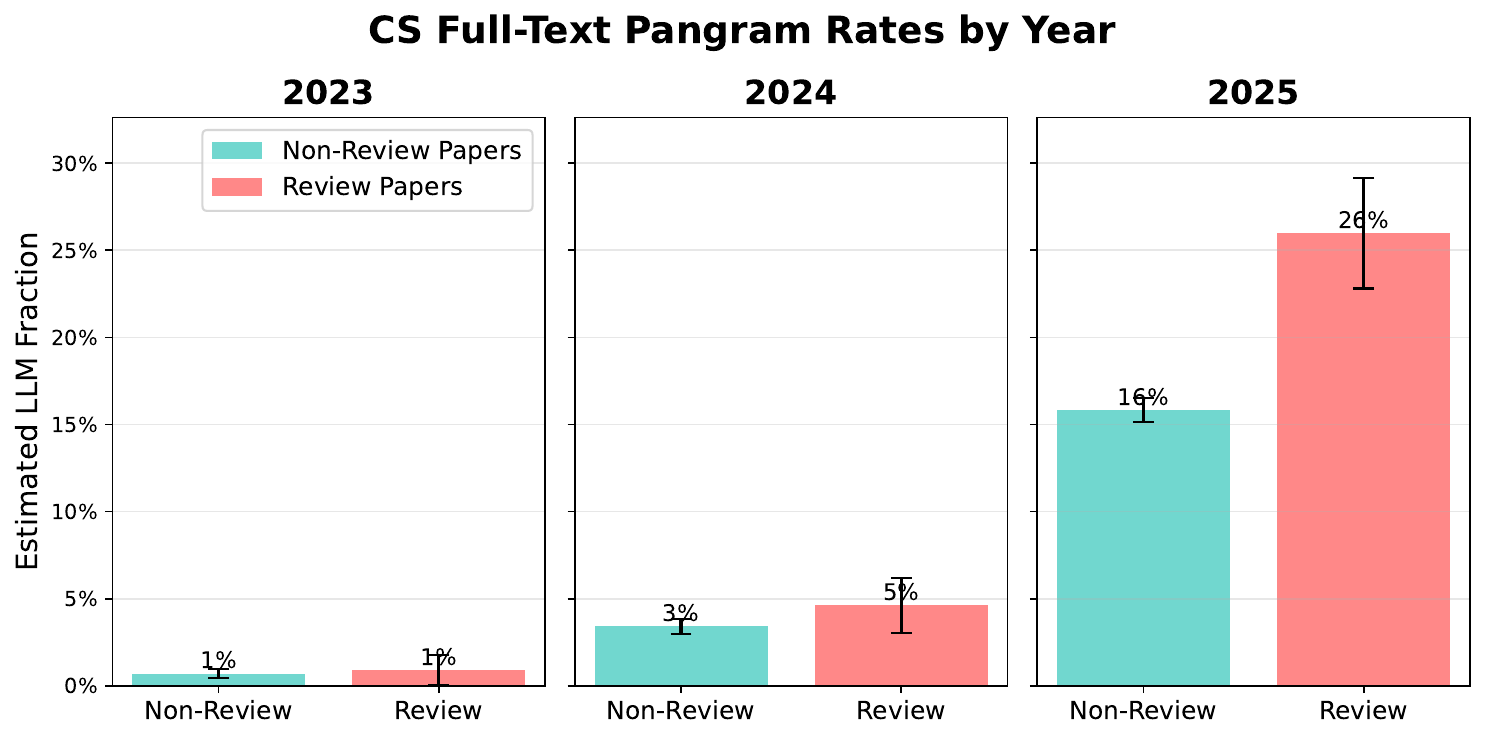}
\caption{LLM-generated papers estimates by year in CS (\textit{arxiv-domains}) using Pangram, when considering the entire text. While the percentages drop by about half compared to the abstract estimates, the yearly upward trend, as well as the review vs. non-review gap persist.}
\label{fig:full-text-pangram-rates}

\end{figure}

So far, we have focused on classifying papers into LLM-generated vs. human-generated (and review vs. non-review) based on the title and abstracts alone.
While this decision may seem naive at first, it is a deliberate one.
First, fully LLM-generated papers also include the abstract, and as such, while we may overestimate the usage of LLMs in paper writing, we are less likely to miss any.
Second, and perhaps most important for the context of this work, is that arXiv is basing its initial automated decisions for the types of papers (review vs. non-review) on the title and abstract alone.\footnote{In conversations with a contact internal to arXiv, we were told that the initial filter mechanism is done based on the title and abstract alone. Flagged papers are then passed to a human moderator for a final decision.} %
As such, the implications based on our results reflect the new arXiv policy.

To validate that this simplification produces reliable results, we extend our analysis and quantify LLM-generated content using the entire paper, instead of solely the title and abstract.
In this experiment, we use the post-LLM era CS papers that Pangram classified as LLM-generated. 
We successfully downloaded and extracted the full text from 430 of the 437 papers from that period.
Using Pangram, we find that 57.9\% of the full texts were predicted as LLM-generated.
Split by paper type, we find that 216 out of 365 non-review papers (59.2\%) are classified as LLM-generated, compared to 33 out of 65 of review papers (50.8\%) (again, of the papers already classified by Pangram as being generated).
These results suggest that review papers tend to use LLMs for generating abstracts over the entire paper more often then non-review papers.
Using the full texts, we also confirm the trends reported in \S\ref{subsection:upward-trends} and find that overall, review papers are more likely to be predicted as LLM-generated, and these rates are increasing over time (see Figure \ref{fig:full-text-pangram-rates}). For instance, 26\% of the CS papers in 2025 were classified as review, vs. 16\% of the \regular{} papers.
These results provide additional support for questioning the new arXiv policy.

\section{Discussion} \label{sec:discussion}

Our findings provide quantitative evidence that review papers on arXiv contain higher proportions of LLM-generated content than non-review papers, \textit{seemings} to support arXiv's stated rationale for their policy change. However, further analysis of the results reveals crucial nuances and unintended effects that warrant reconsideration of this policy.

\paragraph{Magnitude of the Difference.}
Over the past three years, we estimate the use of LLMs in CS papers to be 21.4\% in review papers compared to 14.0\% in \regular{} papers. Considering the per-year analysis, this gap increases considerably (e.g., a 6.7\% difference in 2023 to 9.3\% in 2025 using the Alpha estimate, and 4.7\% in 2023 to 20.0\% using Pangram). 
Crucially, however, \textbf{when considering the number of papers from each type, the worry about LLM usage in reviews fades completely.} We estimate the number of \regular{} papers to be 141K compared to 12K review papers in CS in 2025.
Thus, even though the difference in review percentages is substantial in 2025 (43.3\% vs. 23.3\% using Pangram), \textbf{the number of review papers generated by LLMs is estimated to be 4,783 compared to 26,801 for \regular{} papers} (Figures \ref{fig:pangram-estimates-cs-physics} \& \ref{fig:pangram-estimates-math-stats}, and Tables \ref{tab:papers-computer-science-stats-combined}-\ref{tab:papers-statistics-stats-combined} in Appendix \ref{app:review-number-estimates}.).
As such, the arXiv ban on unpublished review papers makes little sense, if the goal is to save moderators time and energy, as the overall burden from review papers is tiny in comparison to the very large number of \regular{} papers using AI in paper writing.

\paragraph{Disciplinary Variations Matter.}
The cross-domain analysis (Figures~\ref{fig:adjusted-alpha-category-year} and \ref{fig:adjusted-alpha-category}) reveals that AI adoption patterns vary considerably across domains. CS shows the highest prevalence, unsurprising given the domain's focus on AI technology and early access to these tools.
However, the review versus non-review paper gap persists across most examined domains, suggesting this is a general phenomenon rather than one specific to CS.

The temporal patterns are particularly revealing. The sharp increase following ChatGPT's release in late 2022 is evident across all domains, but the rate of adoption differs. CS shows earlier and faster adoption, while traditional domains like Mathematics show more gradual increases. This suggests that domain-specific norms and tool familiarity might influence AI adoption rates.

Our subcategory analysis within CS (\S\ref{subsec:subcategories}) further demonstrates that even within a single domain, AI adoption varies dramatically. Certain subfields (Cryptography and Security, AI) show substantially higher adoption rates than others (Computation and Language, Information Retrieval), and the review-non-review gap manifests differently across subcategories. 
This fine-grained heterogeneity suggests that blanket policies applied uniformly across all CS subcategories may be overly broad.

The validation of these subcategory patterns through Pangram analysis adds an important dimension to understanding AI adoption across CS communities. 
Crucially, the interdisciplinary subcategory Computers \& Society displays much higher rates of review papers, meaning that \textbf{under the new policy, nearly 50\% of Computers \& Society papers would be censored by arXiv, in contrast to only 3\% of Computer Vision papers.}
Our analysis of topics from OpenAlex reemphasizes this point, as we found that \textbf{papers about education and safety (timely and important topics as society grapples with the effects of AI) will be censored at much higher rates than papers focused on engineering topics}.

\paragraph{Implications for Policy.}
Our results raise serious questions about arXiv's decision to restrict review papers. Considering the raw counts, we would be better off banning non-review papers if our goal is to save moderators time.
And if the concern is LLM-generated content quality, should the policy focus solely on review papers when \regular{} papers also show substantial (albeit lower) AI content? 
More broadly, given the sharp increases in AI usage across fields, we need a general policy on the usage of AI in academic writing, beyond CS. 

An alternative interpretation is that arXiv's policy targets not the use of LLMs in itself but the \emph{type} of content that CS arXiv moderators perceive as ``easiest'' to generate with current AI systems—synthesis and summarization. 
While we do not investigate such claims in this paper, as research improves in developing automated scientists \citep{lu2024ai,yamada2025ai}, the ease of fully generating AI ``regular'' papers decreases substantially, making this assumption questionable.

\paragraph{The Future of Scientific Publishing.}
Our cross-disciplinary findings suggest that LLM adoption in academic writing is not a temporary phenomenon but an ongoing trend that varies across fields. As LLM writing tools improve and become more integrated into more research workflows, we may see convergence in adoption rates across disciplines. 
On the other hand,  as academic communities develop more established norms (ethical or otherwise), we may see further divergence.
Finally, these patterns may evolve as both LLMs and detection models improve and as authors develop more sophisticated usage strategies.

Publishers and preprint servers will need to develop clear, well-thought-out, and reasonable policies that account for disciplinary differences while maintaining scientific integrity and considering the development and incorporation of AI-assisted tools. The heterogeneity revealed by our subcategory analyses suggests that blanket restrictions may be overly broad, while the substantial LLM content in \regular{} papers indicates that concerns should extend beyond review papers.

\section{Conclusions}

Our study evaluates the evidence behind arXiv's 2025 restriction on review papers in the CS domain. Across two independent LLM detectors (Alpha and Pangram) and an accurate review vs.
\regular{} classifier, we find that review papers consistently contain higher proportions of LLM-generated content. However, the much larger volume of \regular{} papers means that they account for the vast majority of LLM-generated manuscripts on arXiv, challenging the premise of a review-only ban.

Temporal and disciplinary analyses reveal a sharp post-ChatGPT rise in AI-assisted writing that spans CS, Physics, Mathematics, and Statistics, with heterogeneous uptake across CS subfields. These patterns suggest that LLM-assisted writing is a broad, accelerating shift rather than a CS or review-specific anomaly. Blanket restrictions on review papers risk missing most LLM-generated content while disproportionately affecting communities where synthesis and qualitative work are core scholarship.
We advocate for evidence-based moderation that combines transparent monitoring with clearer guidance on acceptable LLM assistance.

\section*{Limitations}

Our study has several important limitations that should be considered when interpreting the results.

\paragraph{Detection Method Limitations.}
The Alpha estimator, while powerful for group-level estimation, relies on distributional assumptions about LLM-generated text. The method may be sensitive to changes in the LLM used for generation, the data used to estimate alpha, or post-editing by human authors. Our pre-LLM baseline validation suggests the Alpha estimator has differential false positive rates for review and regular papers, which we address through type-specific Rogan-Gladen adjustment. However, this assumes the false positive pattern remains stable over time and does not interact with actual AI use. On the other hand, we detected no false positives with the Pangram detector.

\paragraph{Classification Accuracy.}
Our review vs. non-review paper classifier achieves a high classification score (92.0\% F1), but misclassifications could still bias our results. We validated the classifier on held-out data, but some borderline cases (e.g., tutorial papers, hybrid papers with both novel contributions and community norms) may be ambiguous.
In general, our annotations reflect an attempt at a ``typical'' CS view of what a ``review'' paper might represent, as this is likely the stance of many arXiv moderators, rather than trying to represent the many epistemic views of different subfields.

\paragraph{Interdisciplinary Papers.}
Papers in interdisciplinary areas of CS often blend methods and place greater emphasis on position-taking and literature synthesis, even in empirical work. 
A single paper might present an argument about how the field should develop, review existing literature to support that argument, and include empirical analysis to substantiate its claims. 
These functions are not mutually exclusive; they can coexist within a paper, which may be published in general venues rather than isolated in, e.g., ``position paper tracks.''
While this is a limitation for our work---we had to rely on a simple definition of review papers for our annotations---it is an even larger limitation for the arXiv policy that will need to moderate, across a large number of papers, which papers are reviews.

\paragraph{Data Sampling}
A key limitation of our work is that our data is restricted to papers \textit{posted} to arXiv rather than \textit{submitted} to arXiv. 
This data is not publicly available, and our attempts to request this data were not successful.
We therefore cannot provide quantitative evidence of the number of papers or distribution of papers that are submitted to arXiv but then flagged by moderators to never be posted.
This likely means that our results in this paper are lower bounds for the rates of generated papers, as we believe some level of moderation has been in place to detect these.
But we have no knowledge of any preceding policy about review papers until the current proposal.

\paragraph{Ethical Considerations.}
Labeling content as ``LLM-generated'' can be inaccurate and carries potential stigma. 
We emphasize that AI assistance in writing is not inherently problematic—many legitimate uses exist, including improving clarity for non-native English speakers. 
A limitation of current LLM-detection methods is that they cannot distinguish between different types of LLM usage in writing. 
For instance, we may want to treat differently papers that were fully written by LLMs versus those where LLMs were used to assist with writing.
Our goal is to provide empirical evidence for policy discussions, not to judge individual authors or papers.

\section*{Acknowledgments}

We thank Paul Ginsparg for his insights about arXiv and providing us with useful insights about the internal processes behind arXiv, as well as Thomas Dietterich for the meaningful discussions.
We would also like to thank Pangram for providing us with free credits for their platform.

\bibliography{custom}

\appendix

\clearpage

\section{Rates Vary Dramatically By Discipline: Ellaboration}
\label{app:rates-expanded}

\begin{figure}[ht]
\centering
\begin{subfigure}[b]{1.\columnwidth}
\centering
\includegraphics[width=\textwidth]{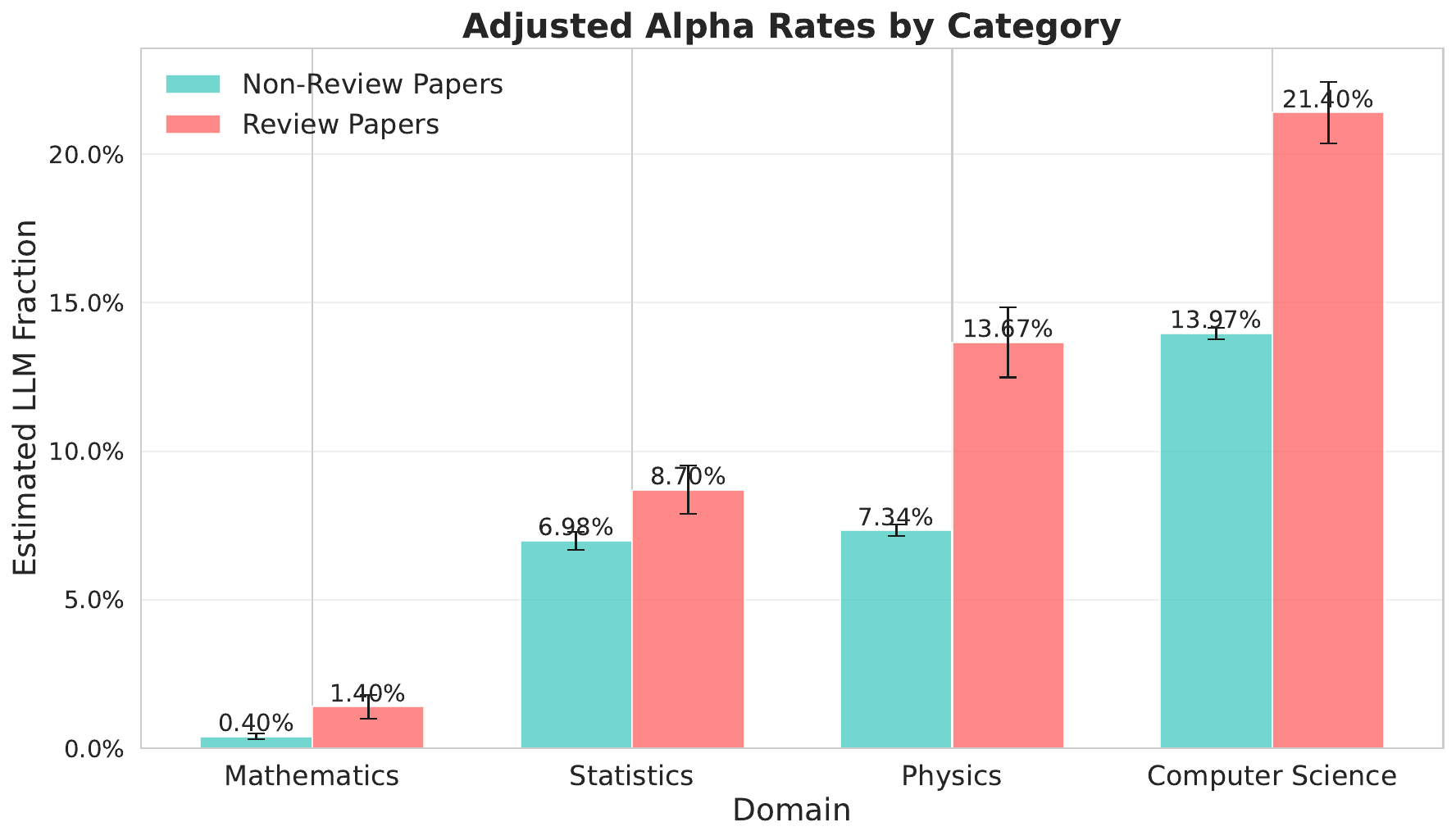}
\caption{Alpha estimates.}
\label{fig:adjusted-alpha-category-alpha}
\end{subfigure}

\begin{subfigure}[b]{1.\columnwidth}
\centering
\includegraphics[width=\textwidth]{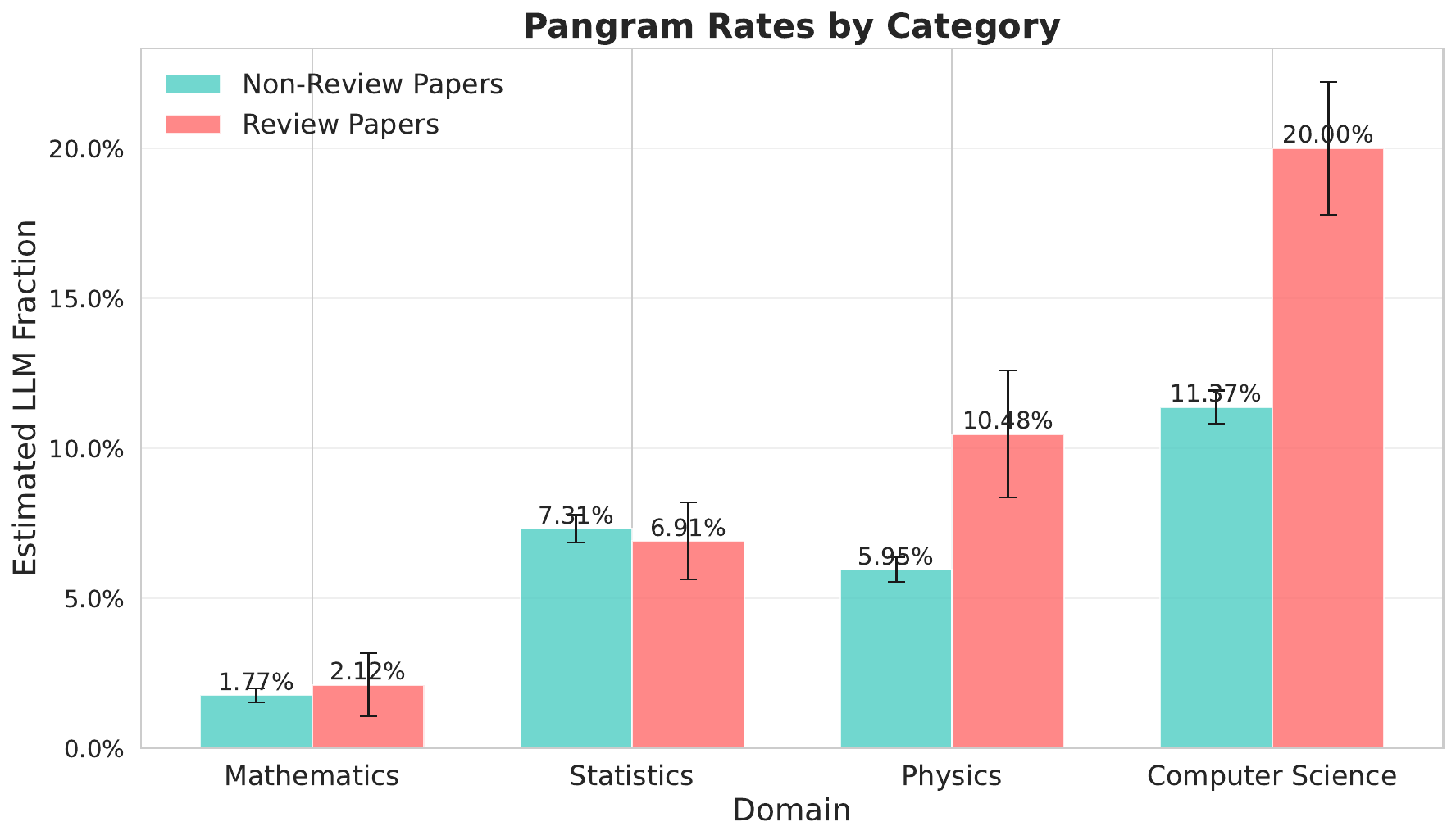}
\caption{Pangram ratios.}
\label{fig:adjusted-alpha-category-pangram}
\end{subfigure}
\caption{LLM-generated papers estimates by domain and paper type on the \textit{arxiv-domains} dataset using (a) Alpha estimates and (b) Pangram detection ratios. Review papers consistently show higher LLM-generated ratios for CS and Physics, and mixed results for Statistics and Math depending on the method. Results are averaged across the post-LLM years (2023-2025).}
\label{fig:adjusted-alpha-category}
\end{figure}

\paragraph{Computer Science.}
We find that review papers have significantly higher LLM-generated content than \regular{} papers in CS. In the post-LLM period (2023-2025), the Alpha estimator reveals that review papers have $\alpha = 21.40\%$, compared to $\alpha = 13.97\%$ for \regular{} papers. %
Similarly, the Pangram results show a similar trend: $20.00\%$ review papers compare to $11.37\%$ \regular{} papers. Figure \ref{fig:adjusted-alpha-category} showcase the year-aggregated results, whereas Figure \ref{fig:adjusted-alpha-category-year-pangram} shows the results split by year.
However, we also note that while the predicted rates are higher for review papers, overall there are far more non-review papers, and so the total number of non-review papers that are predicted to be generated is much higher than for review papers (Figure \ref{fig:pangram-estimates-cs-physics}).

\paragraph{Domain Variations.}

To understand whether LLM-generated content prevalence varies across scientific disciplines, we analyze papers from four major arXiv categories. %
Figure~\ref{fig:adjusted-alpha-category} presents the Alpha and Pangram estimates for the post-LLM period.

We observe notable domain differences. CS papers show the highest overall LLM-generated content, likely reflecting the field's early adoption of AI tools and researchers' familiarity with LLM technology. The review vs. regular paper gap persists across all disciplines\footnote{Aside from Statistics and Mathematics using Pangram, where we do not observe a statistically significant difference between the paper types.} but varies in magnitude, with some fields showing more pronounced differences than others. This suggests that while the general trend holds across fields, the extent of AI adoption in writing depends on domain norms.

\section{Review Paper Classification Details}
\label{app:review}

\begin{table}[h]
\centering

\begin{tabular}{lccc}
\toprule
\textbf{Model} & \textbf{Precision} & \textbf{Recall} & \textbf{F1} \\
\midrule
Llama 3.3 & 92.6 & 87.7 & 90.1 \\
Gemma 2 & 92.9 & 91.2 & 92.0 \\
GPT-OSS & 93.9 & 80.7 & 86.8 \\
GPT-4o-mini & 92.9 & 91.2 & 92.0 \\
\bottomrule
\end{tabular}
\caption{Review vs. Non-Review paper classification performance of different models on the test set.}
\label{tab:classification-results}
\end{table}

We report the results on the test data ($n=140$) for the four models we consider in Table \ref{tab:classification-results}. Llama 3.3, Gemma 2, GPT-OSS, and GPT-4o-mini achieves F1 scores of 90.1, 92.0, 86.8, and 92.0 respectively. We select the model to use for all our experiments based on the validation set (\S \ref{subsec:review-classification}) which was Gemma 2. While GPT-4o-mini performed on par (both on the validation set and the test set), Gemma 2 has the advantage that it is an open-weights model, and we are able to run it locally.

\paragraph{LLM Prompt}
The following prompt was used to classify papers as review/survey papers versus other types of papers:

\begin{tcolorbox}[colback=gray!10, colframe=gray!50, boxrule=0.5pt, arc=2mm]
\small
You are an expert academic paper classifier. Your task is to classify papers into one of two categories:

\textbf{1. Review/Survey}: Papers that primarily review, survey, synthesize existing research, or present positions/perspectives on a field. These papers:
\begin{itemize}
  \item Systematically review existing literature
  \item Provide comprehensive overviews of a research area
  \item Summarize and compare multiple existing approaches
  \item Present position papers or perspectives on research directions
  \item Often have titles containing words like ``survey'', ``review'', ``overview'', ``primer'', ``position'', ``perspective''
\end{itemize}

\textbf{2. Other}: All other papers including:
\begin{itemize}
  \item Original research papers presenting new methods, experiments, or results
  \item Technical reports
  \item Case studies
  \item Papers presenting new datasets or benchmarks
  \item Papers focused primarily on novel empirical results
\end{itemize}

You must respond with ONLY a JSON object in this exact format:

\texttt{\{"classification": "review", "reasoning": "brief explanation"\}}

OR

\texttt{\{"classification": "other", "reasoning": "brief explanation"\}}

The classification field must be either ``review'' or ``other'' (lowercase). Keep the reasoning brief (1-2 sentences).
\end{tcolorbox}

For each paper, the classifier was provided with the paper's title and abstract in the user prompt.
We did not optimize the prompt, as it worked well out of the box.

\section{Dataset Statistics by Year and Category}
\label{app:dataset-stats}

We provide statistics of the two datasets we collected.
Tables \ref{tab:main_categories} and \ref{tab:cs_subcategories} show the number of papers collected for the \textit{arxiv-domains} and \textit{cs-subcategories} datasets, respectively, broken down by year and subcategory.
For each category, we sample up to 500 papers per month from 2020-2025.

\begin{table*}[ht]
    \centering
    \begin{tabular}{lrrrrrrr}
        \toprule
        \textbf{Category} & \textbf{2020} & \textbf{2021} & \textbf{2022} & \textbf{2023} & \textbf{2024} & \textbf{2025} & \textbf{Total} \\
        \midrule
        Computer Science & 6,000 & 6,000 & 6,000 & 6,000 & 6,000 & 6,000 & 36,000 \\
        Mathematics & 6,000 & 6,000 & 6,000 & 6,000 & 6,000 & 6,000 & 36,000 \\
        Physics & 6,000 & 6,000 & 6,000 & 6,000 & 6,000 & 5,963 & 35,963 \\
        Statistics & 2,409 & 2,500 & 2,521 & 2,654 & 3,034 & 3,380 & 16,498 \\
        \midrule
        Total & 20,409 & 20,500 & 20,521 & 20,654 & 21,034 & 21,343 & 124,461 \\
        \bottomrule
    \end{tabular}
    \caption{Number of papers in the \textit{arxiv-domains} data, broken by category and year.}
    \label{tab:main_categories}
\end{table*}
\begin{table*}[ht]
    \centering
    \resizebox*{0.95\textwidth}{!}{
    \begin{tabular}{lrrrrrrr}
        \toprule
        \textbf{Subcategory} & \textbf{2020} & \textbf{2021} & \textbf{2022} & \textbf{2023} & \textbf{2024} & \textbf{2025} & \textbf{Total} \\
        \midrule
        Artificial Intelligence & 727 & 928 & 822 & 1,309 & 2,029 & 3,758 & 9,573 \\
        Computation and Language & 2,698 & 3,283 & 3,379 & 3,951 & 5,496 & 5,516 & 24,323 \\
        Cryptography and Security & 1,103 & 1,198 & 1,252 & 1,449 & 1,752 & 2,226 & 8,980 \\
        Computer Vision and Pattern Recognition & 5,414 & 5,689 & 5,778 & 5,954 & 6,000 & 6,000 & 34,835 \\
        Computers and Society & 584 & 619 & 424 & 590 & 812 & 914 & 3,943 \\
        Human-Computer Interaction & 578 & 742 & 772 & 1,162 & 1,911 & 2,237 & 7,402 \\
        Information Retrieval & 337 & 494 & 539 & 743 & 944 & 1,098 & 4,155 \\
        Machine Learning & 812 & 3,521 & 3,779 & 3,914 & 4,693 & 5,736 & 22,455 \\
        Robotics & 1,175 & 1,633 & 1,915 & 2,648 & 3,534 & 3,762 & 14,667 \\
        Software Engineering & 777 & 1,198 & 980 & 1,308 & 1,575 & 2,073 & 7,911 \\
        \midrule
        Total & 14,205 & 19,305 & 19,640 & 23,028 & 28,746 & 33,320 & 138,244 \\
        \bottomrule
    \end{tabular}
    }
    \caption{Number of papers in the \textit{cs-subcategories} data, broken by category and year.}
    \label{tab:cs_subcategories}
\end{table*}

\begin{figure*}[t]
\centering
\begin{subfigure}[b]{1.\textwidth}
\centering
\includegraphics[width=\textwidth]{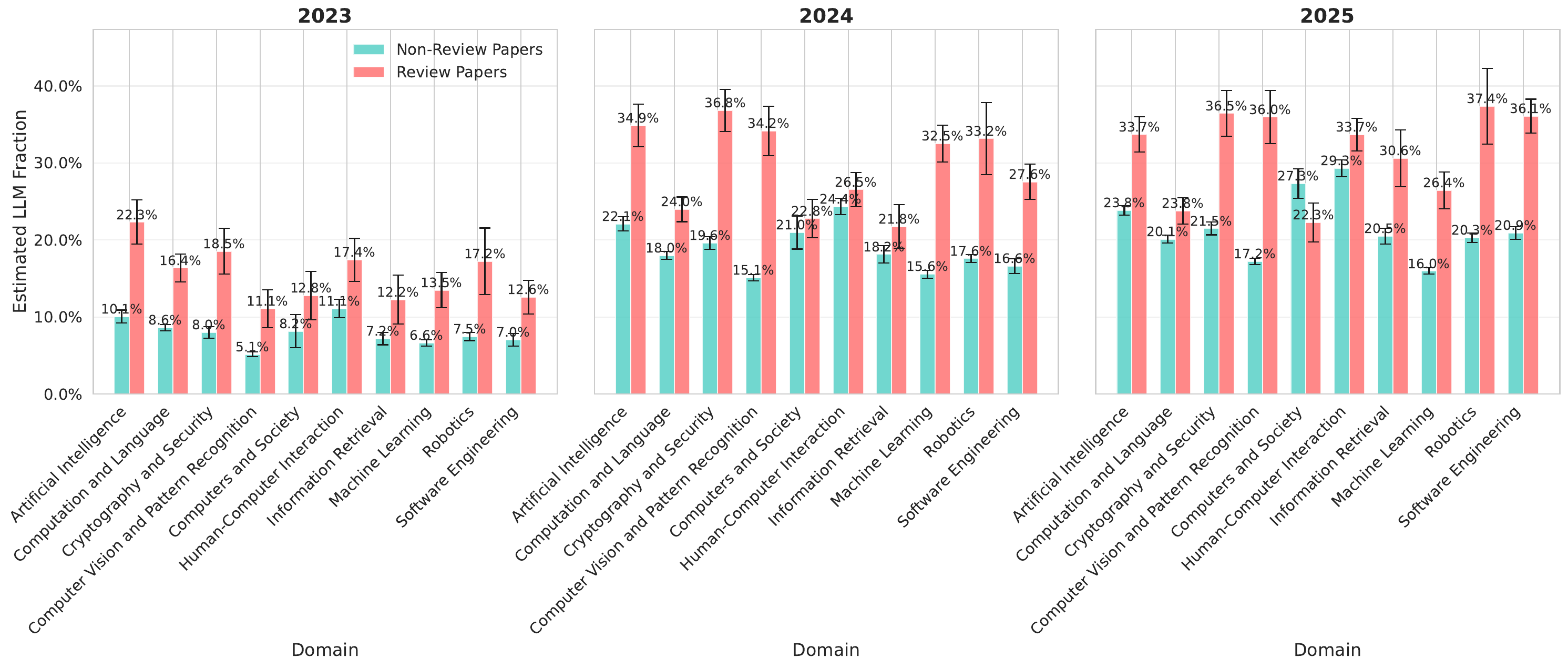}
\caption{Alpha estimator method}
\label{fig:adjusted-alpha-cs-subcat-year-alpha}
\end{subfigure}

\begin{subfigure}[b]{1.\textwidth}
\centering
\includegraphics[width=\textwidth]{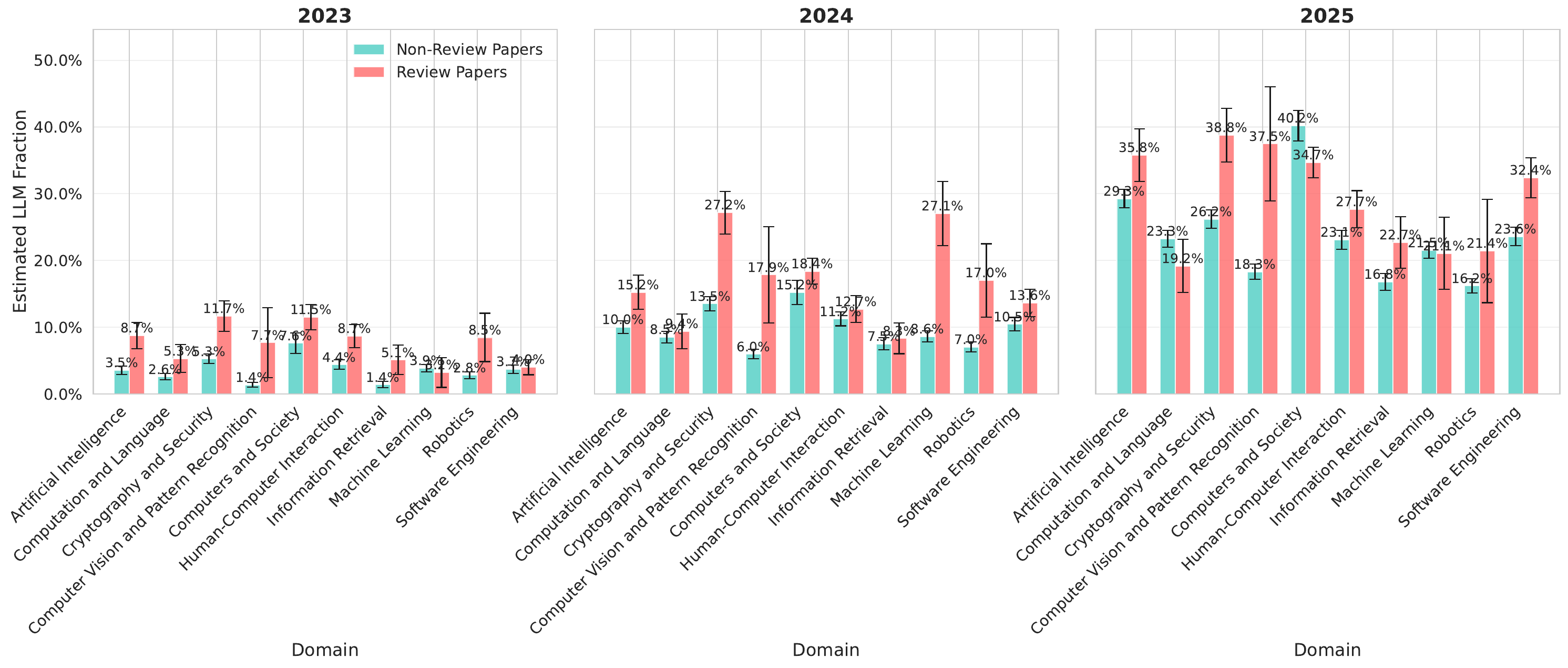}
\caption{Pangram detection method}
\label{fig:adjusted-alpha-cs-subcat-year-pangram}
\end{subfigure}
\caption{LLM-generated papers estimates by Computer Science subcategories in the \textit{cs-subcategories} dataset, paper type, and year using (a) Alpha estimates and (b) Pangram detection ratios. 
}
\label{fig:adjusted-alpha-cs-subcat-year}
\end{figure*}

\section{Estimated review vs. \regular{} papers generated by LLMs}
\label{app:review-number-estimates}

Using our review vs. non-review classifier and the two LLM-paper detection methods we estimate the rates of LLM-generated papers in 2023-2025. By scaling the LLM-detection ratios with the total number of papers per year, we estimate the number of total LLM-generated papers per domain.
We present the results for every domain in Tables \ref{tab:papers-computer-science-stats-combined}-\ref{tab:papers-statistics-stats-combined}

\begin{table}[t]
\centering
\resizebox{\columnwidth}{!}{%
\begin{tabular}{lrrrr}
\toprule
\textbf{Subset / Year} & \multicolumn{2}{c}{\textbf{Review Papers}} & \multicolumn{2}{c}{\textbf{Non-Review Papers}} \\
\cmidrule(lr){2-3} \cmidrule(lr){4-5}
 & \textbf{n.} & \textbf{\%} & \textbf{n.} & \textbf{\%} \\
\midrule
\textbf{All}
 & & & & \\
\quad 2023 & 7,276 & 9.5\% & 69,182 & 90.5\% \\
\quad 2024 & 9,558 & 9.2\% & 94,343 & 90.8\% \\
\quad 2025 & 12,133 & 7.9\% & 141,775 & 92.1\% \\
\addlinespace
\textbf{LLM (Alpha)}
 & & & & \\
\quad 2023 & 937 & 17.9\% & 4,301 & 82.1\% \\
\quad 2024 & 2,298 & 12.5\% & 16,060 & 87.5\% \\
\quad 2025 & 3,420 & 11.3\% & 26,736 & 88.7\% \\
\addlinespace
\textbf{LLM (Pangram)}
 & & & & \\
\quad 2023 & 357 & 18.2\% & 1,607 & 81.8\% \\
\quad 2024 & 868 & 12.8\% & 5,917 & 87.2\% \\
\quad 2025 & 4,783 & 15.1\% & 26,801 & 84.9\% \\
\bottomrule
\end{tabular}}
\caption{Comparison of \textbf{Computer Science} paper estimates by estimation method and year on the \textit{arxiv-domains} dataset. For each subset, we show the estimated review vs. regular papers for each year (2023-2025). }
\label{tab:papers-computer-science-stats-combined}
\end{table}
\begin{table}[ht]
\centering
\resizebox{\columnwidth}{!}{%
\begin{tabular}{lrrrr}
\toprule
\textbf{Subset / Year} & \multicolumn{2}{c}{\textbf{Review Papers}} & \multicolumn{2}{c}{\textbf{Non-Review Papers}} \\
\cmidrule(lr){2-3} \cmidrule(lr){4-5}
 & \textbf{n.} & \textbf{\%} & \textbf{n.} & \textbf{\%} \\
\midrule
\textbf{All}
 & & & & \\
\quad 2023 & 2,038 & 5.7\% & 33,830 & 94.3\% \\
\quad 2024 & 2,524 & 5.9\% & 40,509 & 94.1\% \\
\quad 2025 & 2,889 & 5.0\% & 55,292 & 95.0\% \\
\addlinespace
\textbf{LLM (Alpha)}
 & & & & \\
\quad 2023 & 12 & 15.2\% & 67 & 84.8\% \\
\quad 2024 & 40 & 16.5\% & 202 & 83.5\% \\
\quad 2025 & 66 & 16.6\% & 331 & 83.4\% \\
\addlinespace
\textbf{LLM (Pangram)}
 & & & & \\
\quad 2023 & 0 & 0.0\% & 91 & 100.0\% \\
\quad 2024 & 35 & 10.6\% & 294 & 89.4\% \\
\quad 2025 & 163 & 9.3\% & 1,586 & 90.7\% \\
\bottomrule
\end{tabular}}
\caption{Comparison of \textbf{Mathematics} paper estimates by estimation method and year on the \textit{arxiv-domains} dataset. For each subset, we show the estimated review vs. regular papers for each year (2023-2025). }
\label{tab:papers-mathematics-stats-combined}
\end{table}
\begin{table}[ht]
\centering
\resizebox{\columnwidth}{!}{%
\begin{tabular}{lrrrr}
\toprule
\textbf{Subset / Year} & \multicolumn{2}{c}{\textbf{Review Papers}} & \multicolumn{2}{c}{\textbf{Non-Review Papers}} \\
\cmidrule(lr){2-3} \cmidrule(lr){4-5}
 & \textbf{n.} & \textbf{\%} & \textbf{n.} & \textbf{\%} \\
\midrule
\textbf{All}
 & & & & \\
\quad 2023 & 1,124 & 5.8\% & 18,368 & 94.2\% \\
\quad 2024 & 1,156 & 5.6\% & 19,314 & 94.4\% \\
\quad 2025 & 1,425 & 5.3\% & 25,479 & 94.7\% \\
\addlinespace
\textbf{LLM (Alpha)}
 & & & & \\
\quad 2023 & 60 & 9.1\% & 599 & 90.9\% \\
\quad 2024 & 190 & 11.1\% & 1,516 & 88.9\% \\
\quad 2025 & 277 & 8.9\% & 2,831 & 91.1\% \\
\addlinespace
\textbf{LLM (Pangram)}
 & & & & \\
\quad 2023 & 15 & 8.4\% & 163 & 91.6\% \\
\quad 2024 & 140 & 19.0\% & 597 & 81.0\% \\
\quad 2025 & 240 & 8.7\% & 2,516 & 91.3\% \\
\bottomrule
\end{tabular}}
\caption{Comparison of \textbf{Physics} paper estimates by estimation method and year on the \textit{arxiv-domains} dataset. For each subset, we show the estimated review vs. regular papers for each year (2023-2025). }
\label{tab:papers-physics-stats-combined}
\end{table}
\begin{table}[ht]
\centering
\resizebox{\columnwidth}{!}{%
\begin{tabular}{lrrrr}
\toprule
\textbf{Subset / Year} & \multicolumn{2}{c}{\textbf{Review Papers}} & \multicolumn{2}{c}{\textbf{Non-Review Papers}} \\
\cmidrule(lr){2-3} \cmidrule(lr){4-5}
 & \textbf{n.} & \textbf{\%} & \textbf{n.} & \textbf{\%} \\
\midrule
\textbf{All}
 & & & & \\
\quad 2023 & 401 & 12.5\% & 2,801 & 87.5\% \\
\quad 2024 & 407 & 10.6\% & 3,419 & 89.4\% \\
\quad 2025 & 568 & 9.6\% & 5,358 & 90.4\% \\
\addlinespace
\textbf{LLM (Alpha)}
 & & & & \\
\quad 2023 & 17 & 16.3\% & 87 & 83.7\% \\
\quad 2024 & 28 & 9.6\% & 263 & 90.4\% \\
\quad 2025 & 84 & 14.4\% & 499 & 85.6\% \\
\addlinespace
\textbf{LLM (Pangram)}
 & & & & \\
\quad 2023 & 2 & 4.5\% & 42 & 95.5\% \\
\quad 2024 & 24 & 17.6\% & 112 & 82.4\% \\
\quad 2025 & 57 & 8.3\% & 630 & 91.7\% \\
\bottomrule
\end{tabular}}
\caption{Comparison of \textbf{Statistics} paper estimates by estimation method and year on the \textit{arxiv-domains} dataset. For each subset, we show the estimated review vs. regular papers for each year (2023-2025). }
\label{tab:papers-statistics-stats-combined}
\end{table}

\section{Rogan-Gladen Adjustment}
\label{app:rogan-gladen-adjustment}

The Rogan-Gladen adjustment \citep{rogan1978estimating} corrects for biased prevalence estimates.

Given:
\begin{itemize}
  \item $\hat{p}$ = observed (apparent) prevalence from the test
  \item $Se$ = sensitivity (true positive rate) = $1 - \text{FNR}$
  \item $Sp$ = specificity (true negative rate) = $1 - \text{FPR}$
\end{itemize}

Using simple arithmetic, it can be shown that the prevalence is:

\begin{equation}
\hat{p}_{\text{adj}} = \frac{\hat{p} + Sp - 1}{Se + Sp - 1} = \frac{\hat{p} - \text{FPR}}{1 - \text{FPR} - \text{FNR}}
\end{equation}

In our application, we estimate the false positive rate (FPR) from pre-LLM era papers (2020--2022), when large language models like ChatGPT did not exist. We assume the false negative rate (FNR) is negligible or zero, simplifying Equation~1 to:

\begin{equation}
\hat{p}_{\text{adj}} = \frac{\hat{p} - \text{FPR}}{1 - \text{FPR}}
\end{equation}

This adjusted estimate accounts for the baseline false positive rate of our detector, allowing us to better isolate the true prevalence of LLM-generated content in the post-LLM era.

\section{OpenAlex Data \& Results}
\label{appendix-openalex}

\begin{figure*}[t]
\centering
\includegraphics[width=0.60\textwidth]{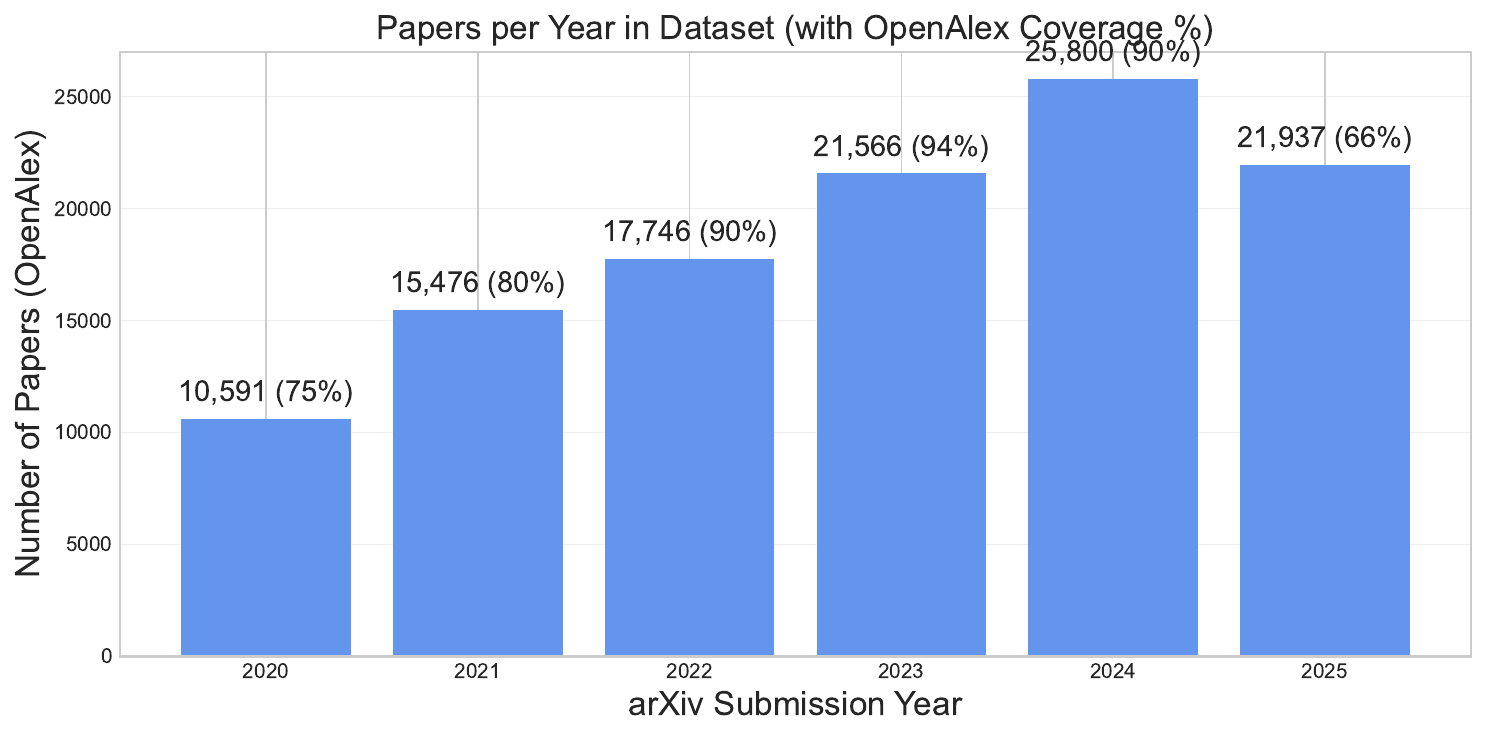}
\caption{The number of papers per year that we were able to match to OpenAlex records. At the top of each column, we show the total number of papers successfully identified as well as the percent of the papers in the original dataset that were identified (the number of papers in the original dataset also increases over time).}
\label{fig:openalex-papers-per-year}
\end{figure*}

We use the OpenAlex API\footnote{\href{https://docs.openalex.org/}{https://docs.openalex.org/}} to enrich the \textit{cs-subcategories} dataset of 138K papers with metadata about the papers, their authors, and the institutional affiliations of the authors.
OpenAlex is a nonprofit online library that contains ``over 250M scholarly works from 250k sources, with extra coverage of humanities, non-English languages, and the Global South.''\footnote{\href{https://openalex.org/}{https://openalex.org/}} 
For each CS paper in the dataset, we use the paper's DOI to collect the paper's topics, keywords, citation count, and authors, as well as each author's affiliation, h-index, number of works, and citation count, and each institution's country and h-index.
Pangram results are available for a subset of these (2.3k–3.6k papers per CS subcategory, 33,413 in total).
We were able to match 82\% of the papers in our sample to OpenAlex records, for a total of 113,116 matched papers.
Match rates rise through 2023 and then decline, with 2025 lowest at 65.8\%, reflecting both OpenAlex's incomplete indexing of recent preprints and the loss of arXiv DOIs from records that have since been published (Fig.~\ref{fig:openalex-papers-per-year}.

We perform a manual audit of the linked data. 
Out of 20 randomly selected papers, we find just two errors in the OpenAlex topics assigned to the papers.
Out of 20 randomly selected authors, we find 9 errors, mostly merging junior authors with ambiguous names with more senior authors and their publications and citation metrics. 
This artificially ``raises'' some junior authors to senior level.

\begin{figure*}[t]
\centering
\includegraphics[width=0.9\textwidth]{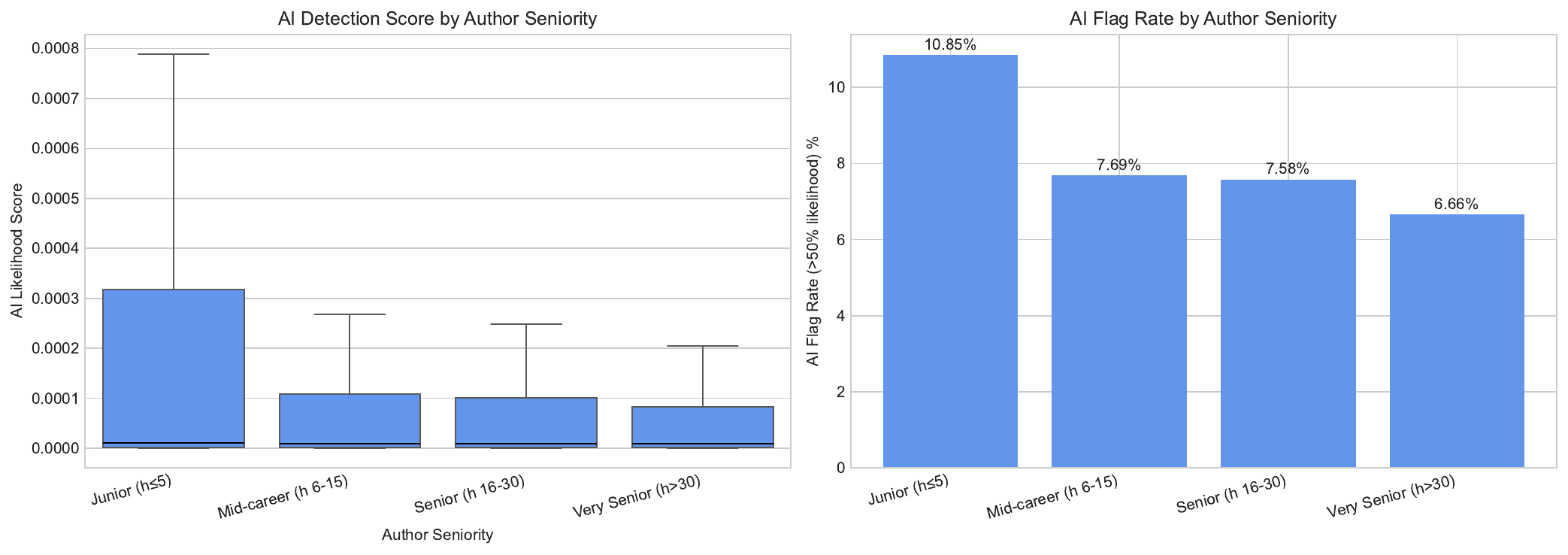}
\caption{Author seniority (measured by h-index) and rates of predicting that those authors' papers are generated. Only papers included in our post-2019 sample are included.}
\label{fig:openalex-seniority}
\end{figure*}

We therefore provide a strong caveat to the following results and caution against over-reliance on OpenAlex author data.
Fig.~\ref{fig:openalex-seniority} shows that junior authors are more likely to have written papers that are flagged as AI-generated by Pangram.
To bin the authors by seniority, we rely on their h-index from OpenAlex.
We use h-index instead of number of works because the works count can be inflated by prolific co-authorship or certain publication patterns, while h-index is more robust to this inflation.
We find that junior authors are more likely to write abstracts that are predicted to be generated by Pangram (11\% of junior authors $h<=5$ compared to 7\% of senior authors $h>30$).
Examining Fig.~\ref{fig:openalex-seniority} , one might worry that the lower rates of generated papers for senior authors could be caused by senior authors writing more papers before these tools were available; however, we are not including all papers by these authors in this analysis, but rather only the papers in our sample, which are all post-2019.
It is true, however, that h-index is measured at the time of scraping, not at the time of writing and posting the paper.
The AI detection score is a continuous value from 0 to 1 representing the probability that a paper was AI-generated, as returned by the Pangram API, while the AI flag rate is a binary yes/no flag derived from the score ($>0.5$).

\begin{figure*}[t]
\centering
\includegraphics[width=0.9\textwidth]{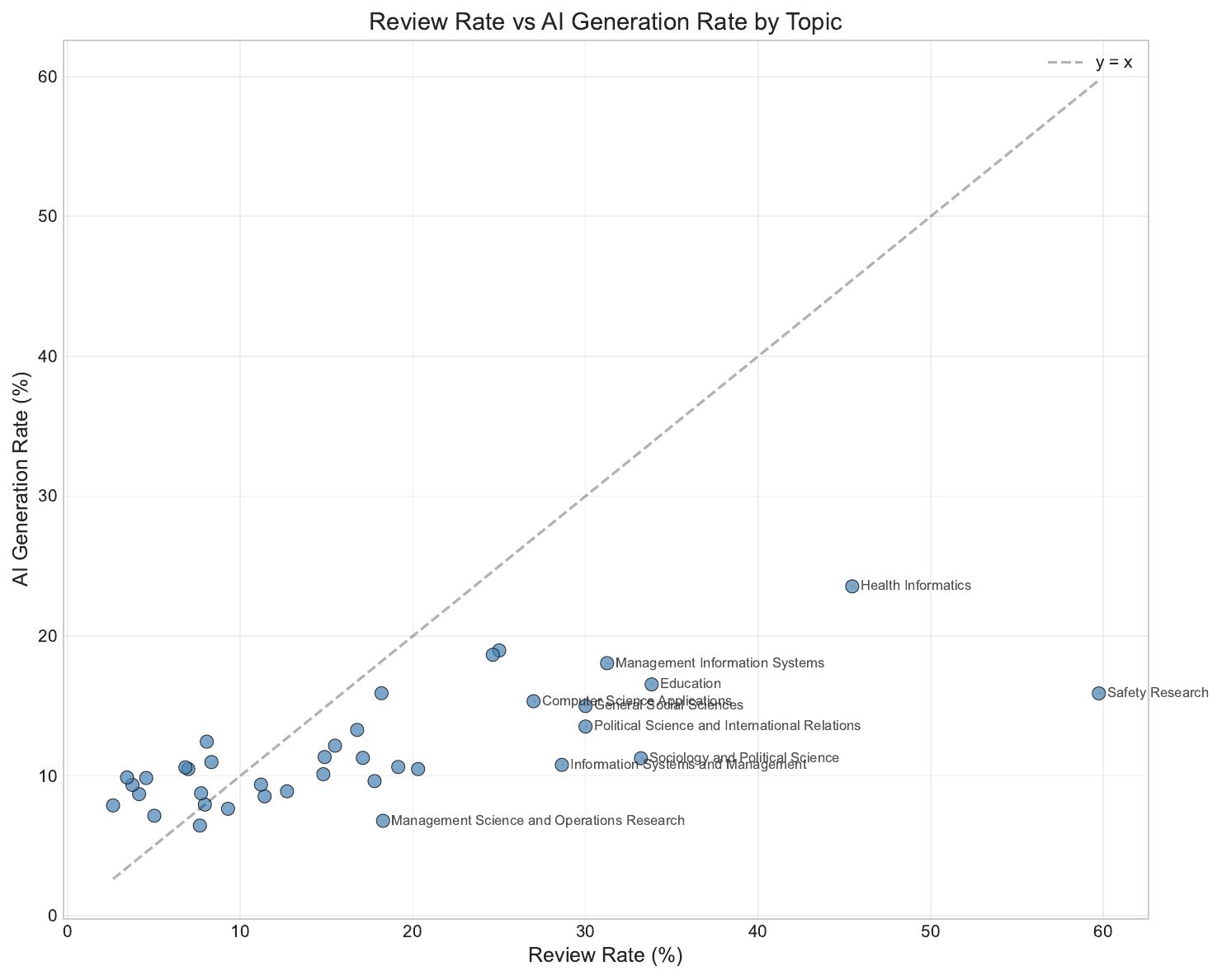}
\caption{Paper topics from OpenAlex according to their predicted review and generation rates.}
\label{fig:openalex-topics-scatter}
\end{figure*}

Finally, in Figure \ref{fig:openalex-topics-scatter}, we show the same topics and rates as in Table \ref{tabel:openalex-topics} but in a scatter plot that emphasizes outliers (like the high-review topics shown before) and shows that the rates of review papers do \textit{not} rise proportionally to the rates of generated papers, despite their overall correlation. 

\section{Additional Details}

\subsection{Data Annotation}
Two authors of this paper performed the data annotation described in \ref{subsec:review-classification}, by the paper's definition of review papers, and used binary labels (review vs. \regular{}) to annotate papers based on the title and abstract alone.

\subsection{AI Assistant Usage}
We used Claude Code to assist in writing the code for this paper, which we manually validated and edited.
We also used AI assistants to help with editing the paper.

\begin{figure}[t]
\centering
\includegraphics[width=0.95\columnwidth]{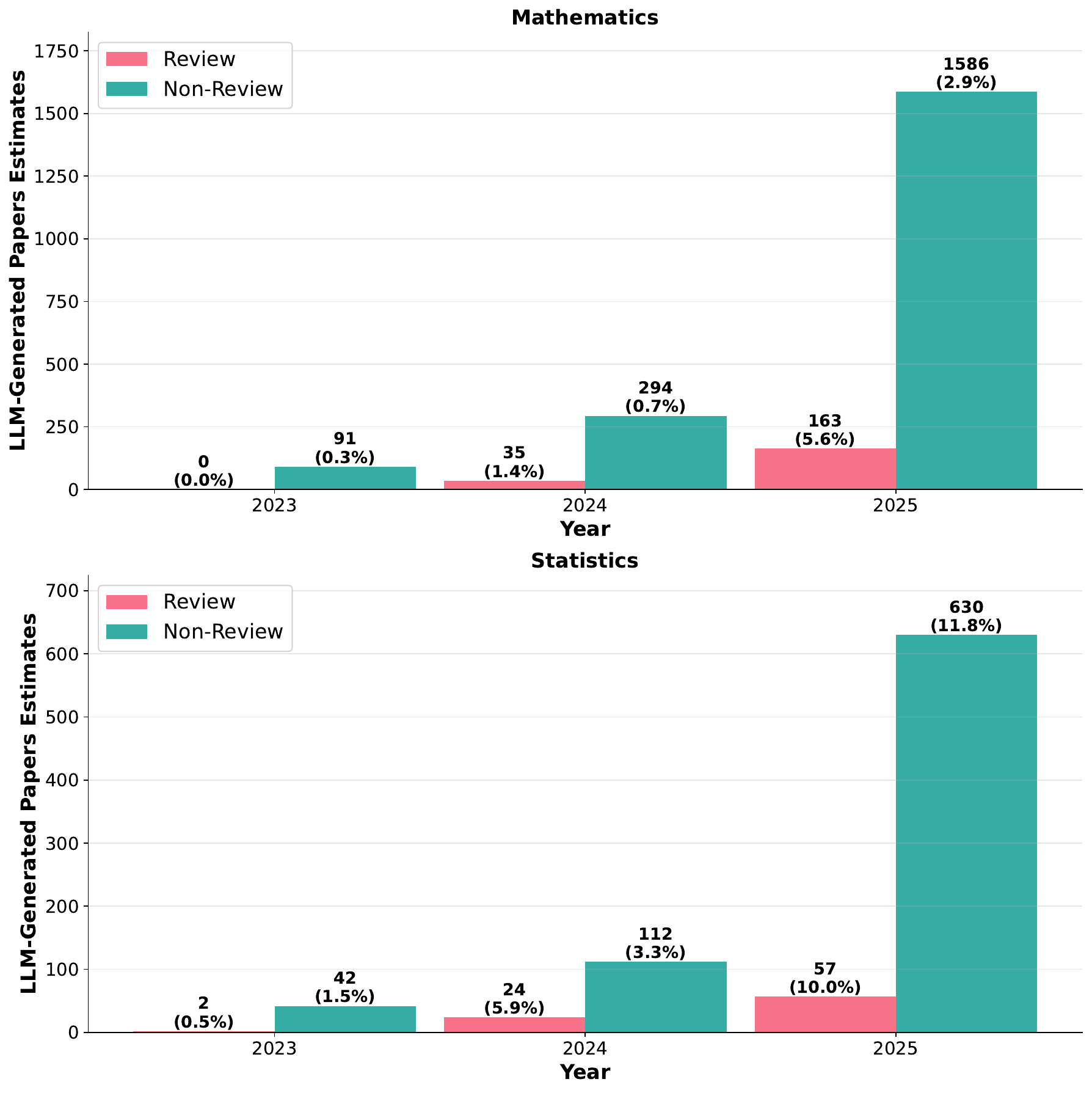}
\caption{LLM-generated paper estimates using Pangram detection of Mathematics and Statistics for review vs. non-review papers on the \textit{arxiv-domains} dataset. Notably, while review papers show higher percentages of LLM-generated content (in parentheses), the absolute number of non-review papers detected as LLM-generated is substantially larger in all domains, suggesting that restricting review papers alone may not address the core concern about LLM-generated content in academic publishing.}
\label{fig:pangram-estimates-math-stats}
\end{figure}

\end{document}